\documentclass[aps,prb,nofootinbib,twocolumn,amsmath,amssymb,superscriptaddress,notitlepage]{revtex4-2}
\usepackage{latexsym}
\usepackage{graphicx}
\usepackage{times,psfrag,subfigure}
\usepackage{enumerate}
\usepackage{amsmath}
\usepackage{amsthm}
\usepackage{dsfont}
\usepackage{dcolumn}
\usepackage{bm,bbm}
\usepackage{wasysym}
\usepackage{color}
\usepackage[dvipsnames]{xcolor}
\usepackage[breaklinks,colorlinks,allcolors=blue]{hyperref}
\usepackage{latexsym,amsmath,amssymb,bm,euscript}
\usepackage{textcomp}
\usepackage{tabularx}
\usepackage{multirow}
\usepackage{setspace}
\usepackage{ctable}
\usepackage{sidecap}
\usepackage{placeins}
\usepackage{threeparttable}
\usepackage{braket}
\usepackage{multirow}
\usepackage{comment}
\usepackage{float}
\usepackage{framed}

\usepackage[all]{xy}
\newcommand{\drawgenerator}[8]{%
\xymatrix@!0{%
& #8 \ar@{-}[ld]\ar@{.}[dd] \ar@{-}[rr] & & #7 \ar@{-}[ld]  \\%
#1 \ar@{-}[rr] \ar@{-}[dd] &  & #2 \ar@{-}[dd] &            \\%
& #6 \ar@{.}[ld] &  & #5 \ar@{-}[uu] \ar@{.}[ll]       \\%
#3 \ar@{-}[rr] &  & #4 \ar@{-}[ru]                       %
}%
}

\usepackage{ulem}

\newtheorem{theorem}{Theorem}[section]
\newtheorem{definition}{Definition}[section]

\DeclareMathOperator\coker{coker}
\DeclareMathOperator\im{im}
\DeclareMathOperator\lcm{lcm}

\let \oldbm \bm
\renewcommand{\vec}[1]{\oldbm{#1}}

\def\B{\mathcal{B}}
\def\A{\mathcal{A}}

\def\R{\mathcal{R}}

\def\G{\mathcal{G}}
\def\M{\mathcal{M}}

\def\1{\mathbbm{1}}

\def\F{\mathbb{F}}

\def\Z{\mathbb{Z}}

\makeatletter
\renewcommand*\env@matrix[1][*\c@MaxMatrixCols c]{%
  \hskip -\arraycolsep
  \let\@ifnextchar\new@ifnextchar
  \array{#1}}
  
\makeatother

\allowdisplaybreaks

\graphicspath{{figures/}}
\begin{document}

\title{Entanglement renormalization of fractonic anisotropic $\Z_N$ Laplacian models
}

\author{Yuan Xue}
\email{yuan\_xue@utexas.edu}
\affiliation{Department of Physics, The University of Texas at Austin, Austin, TX 78712, USA}

\author{Pranay Gorantla}
\email{gorantla@uchicago.edu}
\affiliation{Kadanoff Center for Theoretical Physics \& Enrico Fermi Institute, University of Chicago, Chicago, IL 60637, USA}

\author{Zhu-Xi Luo}
\email{zhuxi\_luo@gatech.edu}
\affiliation{School of Physics, Georgia Institute of Technology, Atlanta, GA 30332, USA}
\affiliation{Department of Physics, Harvard University, Cambridge, MA 02138, USA}

\date{\today}

\begin{abstract}
    Gapped fracton phases constitute a new class of quantum states of matter which connects to topological orders but does not fit easily into existing paradigms. They host unconventional features such as sub-extensive and robust ground state degeneracies as well as sensitivity to lattice geometry. We investigate the anisotropic $\Z_N$ Laplacian model \cite{PhysRevB.107.125121} which can describe a family of fracton phases defined on arbitrary graphs. Focusing on representative geometries where the 3D lattices are extensions of 2D square, triangular, honeycomb and Kagome lattices into the third dimension, we study their ground state degeneracies and mobility of excitations, and examine their entanglement renormalization group (ERG) flows. All models show bifurcating behaviors under ERG but have distinct ERG flows sensitive to both $N$ and lattice geometry. 
    In particular, we show that the anisotropic $\Z_N$ Laplacian models defined on the extensions of triangular and honeycomb lattices are equivalent when $N$ is coprime to $3$. We also point out that, in contrast to previous expectations, the model defined on the extension of Kagome lattice is robust against local perturbations if and only if $N$ is coprime to $6$. 
\end{abstract}

\maketitle

\tableofcontents
\section{Introduction\label{Sec: Intro}}
Gapped fracton phases \cite{PhysRevLett.94.040402,PhysRevB.92.235136,PhysRevB.94.235157,PhysRevA.83.042330}  lie at the frontier of our understanding of phases of matter \cite{RevModPhys.96.011001,annurev:/content/journals/10.1146/annurev-conmatphys-031218-013604,doi:10.1142/S0217751X20300033}. While resembling topological phases in many aspects, they do not completely fall into the existing theoretical paradigms: for example, they also host fractionalized excitations, but with restricted mobility; they also exhibit robust ground state degeneracy (GSD) on nontrivial manifolds, but the degeneracy can be sub-extensive and is sensitive to lattice geometry. Even more intriguingly, in the cases usually referred to as type-II fractons such as Haah's cubic code \cite{PhysRevA.83.042330} (also see recent progress such as \cite{PhysRevB.109.075164,PhysRevLett.128.115301}), there exists a sharp fluctuation of the GSD as the system size grows, which renders the definition of phases in the thermodynamic limit challenging.

One important method in modern condensed matter physics to identify phases of matter 
is through the entanglement renormalization (ER) \cite{PhysRevLett.99.220405,PhysRevLett.100.070404}. It is similar to the usual real-space renormalization group (RG) flow \cite{RevModPhys.47.773}, but has an extra step of reducing the short-range entanglement after coarse graining. It has been successfully employed in many systems such as topological phases \cite{PhysRevLett.100.070404,PhysRevB.79.195123}, critical phenomena \cite{PhysRevLett.100.130501,PhysRevA.77.052328,PhysRevB.79.144108,PhysRevA.79.040301} and quantum fields \cite{PhysRevLett.110.100402}.  Progresses using ER to study fracton phases will be reviewed below. 

A general ER of a Hamiltonian $H(a)$ living on a lattice with lattice spacing $a$ can be written as
\begin{equation}
    U H(a)U^\dagger\cong H_1(ca)+H_2(ca)+\dots +H_b(ca),
    \label{Eq: bifurecating_relation}
\end{equation}
where $U$ is a finite-depth local unitary circuit, $b$ is the number of decoupled models and $c$ is the coarse-graining factor. We use $\cong$ to denote that the Hamiltonians on two sides are equivalent up to trivial Hamiltonians whose ground states are product states \cite{PhysRevB.89.075119}.
Conventional fixed-point topological orders or topological quantum liquids (TQL) \cite{PhysRevB.91.125121}, such as toric code \cite{KITAEV20032}, satisfy
\begin{equation}
    UH(a)U^\dagger\cong H(ca). 
    \label{Eq: fix point}
\end{equation}
In two dimensions, it is known that every translational invariant Pauli stabilizer code in 2D can be transformed to copies of 2D toric code via finite-depth local unitary transformations \cite{10.1063/5.0021068,Bombin_2012,Bombín2014}. Therefore, by \eqref{Eq: fix point}, it is equivalent to copies of 2D toric code under ER. In three dimensions, however, the existence of fracton phases in addition to topological phases leads to more possibilities than \eqref{Eq: fix point}. 

\setlength{\tabcolsep}{4pt} 
\renewcommand{\arraystretch}{1.6}
\begin{table*}[t]
    \centering
    \begin{threeparttable}
    \caption{Anisotropic $\Z_N$ Laplacian models defined on 3D extensions of non-square 2D regular lattices  and their entanglement renormalization. The original 3D anisotropic ``parent'' model and its ``child'' model after ER are abbreviated as $\A$ and $\B$ respectively, though we shall only show  ER of their base 2D models in the main text. $\mathrm{STC}$ represents a stack of 2D toric code layers and $\mathrm{FSL}$ \tnote{1} represents the anisotropic fractal spin liquid. The model on the Kagome lattice is not robust for $N=2,3$, so we skip its GSD, mobility and ER. The explicit forms of these models are given in the main text.
    }

    \begin{ruledtabular}
    \begin{tabular}{c c c c c c c}
    $N$ & 2D lattice & Robust? & GSD of $\A$ & Mobility on 2D lattice & ER & ER of $\B$ code \\
    \hline
    2 & Square & Yes & \eqref{Eq: logGSD_Z2_square} & Lineon, fracton & $\A+4\times \mathrm{STC}$ \cite{SHIRLEY2019167922} &  scale-invariant \\
    2 & Triangular/Honeycomb & Yes & \eqref{Eq: log2 GSD tri} & Lineon, fracton & $\A+6\times \mathrm{STC}$ & scale-invariant\\
    3 & Square & Yes & \eqref{Eq: logp GSD sq} & Fracton & $\A+2\times\B$ & $3\times\B$
    \\ 
    3 & Triangular & Yes & \eqref{Eq: logpGSD_tri} & Fracton & $\A+12\times \mathrm{FSL}$  & $ 3\times\mathrm{FSL}$ \\
    3 & Honeycomb & Yes & \eqref{Eq: log3GSD hc} & Fracton & $3\times \A$ & N/A\\
    2, 3 & Kagome & No & - & - & - & -\\
    \end{tabular}
    \end{ruledtabular}
    
    \begin{tablenotes}
    \footnotesize
        \item[1] Here, we dub this model as ``fractal spin liquid" in the sense that it is similar to the Yoshida's fractal spin liquid model introduced in \cite{PhysRevB.88.125122}. However, our model is based on qutrits and defined by the stabilizer matrix in \eqref{Eq: fsl-stab-mat}.
    \end{tablenotes}
    
    \end{threeparttable}
    \label{table: results}
\end{table*} 
 
Fracton phases typically show bifurcating behaviour under ER \cite{PhysRevX.8.031051,PhysRevB.99.115123,10.21468/SciPostPhys.6.4.041,10.21468/SciPostPhys.6.1.015,PhysRevB.89.075119,PhysRevResearch.2.033021,PhysRevB.108.035148,SHIRLEY2019167922}, where they decouple into two independent components \cite{PhysRevLett.112.220502,PhysRevLett.112.240502,PhysRevB.89.235113}. Under a particular coarse-graining, examples of exactly solvable fracton systems have been shown to be either self-bifurcating fixed points satisfying
\begin{equation}
    U H_{SB}(a)U^\dagger\cong n\times H_{SB}(ca),
    \label{Eq: self-bifurcating}
\end{equation}
or quotient fixed points satisfying
\begin{equation}
    U H_{Q}(a)U^\dagger\cong H_Q(ca)+n \times H_{SB}(ca),
    \label{Eq: quotient fixed pt}
\end{equation}
where $n$ is the number of copies. 
While there are many works that study the ER of foliated fracton models  \cite{PhysRevX.8.031051,PhysRevB.99.115123,10.21468/SciPostPhys.6.4.041,10.21468/SciPostPhys.6.1.015,SHIRLEY2019167922}, the ER of type-II fracton phases has has been largely unexplored, with the existing studies \cite{PhysRevA.83.042330,PhysRevResearch.2.033021} focusing on $\mathbb{Z}_2$ qubit cases and simple lattices.

Besides the bifurcating feature, another important property distinguishing fracton orders from TQL is that the former is highly sensitive to lattice geometry. That is, the same fracton model defined on different lattices can have different low-energy properties. This is a consequence of UV/IR mixing in these models \cite{PhysRevB.104.235116}. Recently, a general class of models, which can host fracton phases, have been proposed, where  
the Hamiltonian can be defined on a general spatial graph using the discrete Laplacian operator \cite{PhysRevB.106.195139,PhysRevB.107.125121,10.21468/SciPostPhys.14.5.106,PhysRevB.108.075106,manoj2021arboreal}.
The physical observables are closely related to the graph-theoretic properties: for example, the GSD is related to the complexity of the graph and the global symmetry is identified by the Jacobian group of the graph. 
We will focus on one set of such Laplacian models known as the anisotropic $\Z_N$ Laplacian model \cite{PhysRevB.107.125121,10.21468/SciPostPhys.14.5.106}. On many regular lattices, the GSD of this model is robust against local perturbations. Moreover, it scales sub-extensively in the system size when $N=2$, but exhibits sharp fluctuations when $N$ is an odd prime. 

In this work, we examine the anisotropic $\Z_N$ Laplacian model defined on 3D lattices that are extensions of 2D regular lattices (or graphs), specifically the triangular, honeycomb, and Kagome lattices. More concretely, consider a 2D lattice on the $xy$ plane (or a graph). Its 3D extension (extended lattice) is constructed by stacking the 2D lattices (or graphs) along the third direction, $z$ direction. We sometimes call this the anisotropic direction. The anisotropic model is a 3D model constructed by coupling adjacent 2D models with Ising-type interactions. We analyze the GSD, mobility restrictions of excitations, and robustness for the series of models, and investigate their ERG flows. Below is a brief summary of the results. 
\begin{itemize}
\item GSDs are computed using techniques from commutative algebra \cite{Haah2013}. When $N=2$, the GSD scales linearly with the system size $L$. When $N=p>2$, with $p$ prime\footnote{We always use $p$ to denote an odd prime, while $N$ can be an arbitrary positive integer.}, the GSD fluctuates dramatically with $L$. In both cases, the GSD is sensitive to the geometry of the 2D lattice.
\item Mobilities of point-like excitations are analyzed using the polynomial formalism. In all these models, there always exist excitations (lineons\footnote{A lineon is a point-like excitation that can move only along a line.}) that can move along the anisotropic direction. Adapting the usual terminology, we refer to a model as type-II fracton order if no excitation can move in the $xy$ plane without creating more excitations. The models we study are type-II when $N=p>2$.
\item The $\Z_N$ model on the extended triangular lattice is equivalent to that on the extended honeycomb lattice up to disentangled degrees of freedom when $N$ is coprime to $3$. Models defined on extended triangular and honeycomb lattices are robust, as expected. However, the model defined on the extended Kagome lattice is robust if and only if $N$ is coprime to $6$, again reflecting the sensitivity of the anisotropic Laplacian model to the lattice geometry. 

\item The ER results are summarized in Table \ref{table: results}. All the models we study are bifurcating. When $N=2$, consistent with the items above, the robust models on all geometries studied here bifurcate into themselves and some copies of a scale-invariant topological order. When $N=3$, bifurcating behaviors of these robust models are very similar to that of the cubic code: the resultant model is the original ``parent'' model plus some copies of a different self-bifurcating ``child'' model. In particular, the anisotropic Laplacian model defined on the square or triangular lattice is a quotient fixed point while that defined on the honeycomb lattice is a self-bifurcating fixed point, again an evidence for the geometric sensitivity.
\end{itemize}

The rest of the paper is organized as follows. In Sec. \ref{Section: Polynomial framwork}, we briefly review the polynomial formalism of $\mathbb{Z}_N$ Pauli stabilizer Hamiltonians. In Sec. \ref{Sec: anisotropic}, we review the anisotropic $\Z_N$ Laplacian model on graphs and then specialize to  the extended triangular, honeycomb, and Kagome lattices. The GSD, mobility of excitations, and robustness are discussed. In Sec. \ref{Sec: ER_Laplacian}, we first review the general procedure of ER and then investigate the ER transformations of $\Z_{2,3}$ Laplacian models on the above extended lattices. In Appendix \ref{Appendix: Laplacian_square}, we review the properties of the anisotropic $\Z_N$ Laplacian model on the square lattice. In Appendices \ref{appendix: gsd_ca}, \ref{appendix: mobility_tri_laplacian}, and \ref{appendix: robust-kagome}, we analyze the GSDs, mobilities of excitations, and robustness, respectively, of the models discussed in the main text.

\section{The polynomial framework \label{Section: Polynomial framwork}}

\subsection{The polynomial formalism for stabilizer Hamiltonians}
Any translation-invariant stabilizer Hamiltonian can be expressed conveniently in the polynomial formalism \cite{Haah2013,PhysRevB.88.125122}, which we will use heavily in this work. In this section, we review the basic ideas of this algebraic representation of stabilizer Hamiltonians.

A stabilizer Hamiltonian that includes $q$ types of $\Z_N$ qudits at each site and $t$ types of stabilizers can be written as a $2q\times t$ matrix, denoted as $\A$ or $\B$ in this paper, whose entries are polynomials with coefficients in $\Z_N$, i.e., integers modulo $N$.
Each column corresponds to one of the $t$ stabilizers.
More concretely, working in two spatial dimensions, we have the map
\begin{equation}
    \begin{pmatrix}
        f_1(x,y)\\
        \vdots\\
        f_q(x,y)\\
        \hline
        g_1(x,y)\\
        \vdots\\
        g_q(x,y)
    \end{pmatrix}
    ~\mapsto~ \prod_{i=1}^q S_{Z,i}(f_i) S_{X,i}(g_i),
\end{equation}
where $f_i$'s and $g_i$'s are polynomials, and
\begin{equation}
    \begin{aligned}
        &S_{Z,i}(\sum_{n,m} c_{n,m}~x^n y^m) = \prod_{n,m} Z_{i,(n,m)}^{c_{n,m}},
        \\
        &S_{X,i}(\sum_{n,m} c_{n,m}~x^n y^m) = \prod_{n,m} X_{i,(n,m)}^{c_{n,m}}.
    \end{aligned}
\end{equation}
Here, $Z_{i,(n,m)}$ and $X_{i,(n,m)}$ are the $\Z_N$ Pauli clock and shift matrices, respectively, acting on the $i$-th qudit at site $(n,m)$. They satisfy the algebra $Z^N = X^N = 1$ and $ZX = e^{2\pi i/N} XZ$. Note that each monomial $x^ny^m$ corresponds to the position $(n,m)$ of a qudit and its coefficient $c_{n,m}$ corresponds to the exponent of $Z$ or $X$ in the associated stabilizer term.
Also, the $i$-th row of the first $q$ rows and the $i$-th row of the last $q$ rows represent the parts of the stabilizers that act on the $i$-th qudit. We use $R_i^Z$ to denote the first $q$ rows of the stabilizer matrix, $R_i^X$ to denote the last $q$ rows, and $C_\mu$ to denote the columns. The horizontal line separates the $Z$-type and $X$-type operators in each stabilizer.

Let $\lambda$ be the $2q \times 2q$ matrix given by
\begin{equation}
    \lambda = \begin{pmatrix}[c|c]
        0 & \vec 1_q\\
        \hline-\vec 1_q & 0
    \end{pmatrix},
\end{equation}
where $\vec 1_q$ is the $q\times q$ identity matrix and the $0$'s are zero matrices of appropriate dimension. The fact that all the local terms in a stabilizer Hamiltonian commute with each other is codified into the equation
\begin{equation}\label{Eq: comm-rel}
    \mathcal{A}^\dagger \lambda \mathcal{A} = 0,
\end{equation}
where $\mathcal{A}^\dagger = \overline{ \mathcal{A}}^{\mathsf T}$ and $\overline{\mathcal{A}}$ is the matrix obtained from $\mathcal{A}$ by replacing $x\rightarrow \bar x = x^{-1}$, $y\rightarrow \bar y = y^{-1}$, and so on.

In the following, we restrict to those Hamiltonians whose stabilizers are all either $Z$-type or $X$-type, i.e., each stabilizer involves either only $Z$ operators or only $X$ operators. They are known as CSS codes \cite{doi:10.1098/rspa.1996.0136,PhysRevA.54.1098}. In the polynomial formalism, the stabilizer matrix of such models is of the form
\begin{equation}\label{Eq: css-code}
    \A = \begin{pmatrix}
        \A^Z & 0\\
        \hline 0 & \A^X
    \end{pmatrix}.
\end{equation}

As an example, consider the 2D $\Z_2$ toric code \cite{KITAEV20032} on the square lattice.
Its stabilizers are given by
\begin{eqnarray}\label{TC-stab}
    \begin{array}{c}
        \xymatrix@!0{
            IZ \ar@{-}[r] & II \ar@{-}[d] \\
            ZZ \ar@{-}[u] & ZI \ar@{-}[l]
        } \quad
        \xymatrix@!0{
            IX \ar@{-}[r] & XX \ar@{-}[d] \\
            II \ar@{-}[u] & XI \ar@{-}[l]
        } 
    \end{array},
\end{eqnarray}
where $ZZ$ is shorthand for $Z_1 Z_{2}$ with $Z_1$ acting on the first qubit at the site and $Z_{2}$ acting on the second qubit at the same site; similarly $XI$ stands for $X_1 I_{2}$. In the polynomial formalism, the monomials associated with the four vertices of a plaquette can be taken as
\begin{eqnarray}
    \begin{array}{c}
        \xymatrix@!0{
            y \ar@{-}[r] & xy \ar@{-}[d] \\
            1 \ar@{-}[u] & x \ar@{-}[l]
        }
    \end{array}.
\end{eqnarray}
Therefore, the stabilizer matrix of the toric code is given by
\begin{equation}
    \mathcal{A}_\mathrm{TC}=\begin{pmatrix}
        1+x & 0 \\
        1+y & 0 \\
        \hline
        0 & x+xy \\
        0 & y+xy \\
    \end{pmatrix}.
\end{equation}
where the first (second) column corresponds to the $Z$-type ($X$-type) stabilizer in \eqref{TC-stab}.

Using the fact that $x\Bar{x}=1$, $y\Bar{y}=1$, and by translation invariance, we can transform the above matrix into the equivalent form
\begin{equation}
    \begin{pmatrix}
        1+x & 0 \\
        1+y & 0 \\
        \hline
        0 & 1+\Bar{y} \\
        0 & 1+\Bar{x} \\
    \end{pmatrix},
    \label{Eq: toric_code}
\end{equation}
by multiplying the second column by $\bar{x}\bar{y}$. This operation corresponds to shifting the $X$-stabilizer one step along both $-x$ and $-y$ directions.

\subsection{Elementary symplectic transformations \label{subsection: element_transf}}
In \cite{Haah2013}, it was proved that two equivalent Pauli stabilizer models can be connected by a local unitary, known as a symplectic transformation\footnote{The name is inspired by the fact that the corresponding polynomial matrix is a symplectic matrix that preserves $\lambda$.}. Each symplectic transformation is a composition of some elementary row or column operations \cite{10.1063/5.0021068} that are listed below. Note that while all of them preserve the stabilizer group, and hence the ground states, some of them change the Hamiltonian, and hence the excitations.

There are three kinds of row operations\footnote{Ref. \cite{10.1063/5.0021068} mentions other elementary row operations based on the controlled-phase (CPhase) gate and the Hadamard gate, but we do not use them because they spoil the structure \eqref{Eq: css-code} of the stabilizer matrix.}.
\begin{enumerate}
\item First, we have \begin{equation}
    \mathrm{CNOT}(i,j,f):\; \begin{aligned}
        &R^X_i\mapsto R^X_i+f~R^X_j,
        \\
        &R^Z_j\mapsto R^Z_j-\bar{f}~R^Z_i,
    \end{aligned}
\end{equation}
where $f$ is a polynomial. As indicated, this is implemented by controlled-NOT (CNOT) gates where the target qudits are specified by $i$ and $f$, and the control qudit is the $j$-th qudit.

\item Next, we have
\begin{equation}
    R^X_i\mapsto x^n y^m R^X_i,\qquad R^Z_i\mapsto x^n y^m R^Z_i.
\end{equation}
This corresponds to translating the $i$-th qudit by $(n,m)$ in each stabilizer.

\item And finally, we have
\begin{equation}
    R^X_i\mapsto k R^X_i,\qquad R^Z_i\mapsto r R^Z_i,
\end{equation}
where $k$, $r$ are integers such that $kr = 1 \mod N$. This corresponds to replacing $X_i\rightarrow X_i^k$ and $Z_i \rightarrow Z_i^r$ in each stabilizer. The choice of $k$ and $r$ ensures that the new set of stabilizers still commute with each other. Note that this operation is trivial for qubits.
\end{enumerate}

Similarly, there are three kinds of column operations.
First, we can multiply a column $C_\nu$ by a polynomial $f$ and add it to another column $C_\mu$, i.e.
\begin{equation}
    \mathrm{Col}(\mu,\nu,f):\; C_\mu \mapsto C_\mu + f~ C_\nu.
\end{equation}
Clearly, this amounts to redefining the stabilizer generators. Next, we can translate a stabilizer by $(n,m)$ by multiplying the associated column by $x^n y^m$. And finally, we can multiply a column by an integer that has an inverse modulo $N$. The third operation is trivial for qubits.

\subsection{Anisotropic extension in the polynomial formalism}\label{subsection: anisotropic extension}

Consider a stabilizer Hamiltonian described by the $2q\times t$ matrix
\begin{equation}
    \A = \begin{pmatrix}
        \A^Z_{q\times t}\\
        \hline
        0
    \end{pmatrix},
\end{equation}
where the $0$ in the bottom row is understood to be a zero matrix of size $q\times t$. The form of this matrix implies that all stabilizers are of $Z$-type. Moreover, it satisfies \eqref{Eq: comm-rel} trivially. An example of this type is the 1D Ising model without the transverse-field term, which has the stabilizer matrix
\begin{equation}\label{Ising-stabmat}
    \A_\mathrm{Is}=\begin{pmatrix}
        1+x\\
        \hline
        0
    \end{pmatrix}.
\end{equation}
Such models are commonly referred to as ``classical codes'' (see, for example, \cite{PhysRevB.88.125122}).

When $q=t$, one can define the anisotropic extension of the classical code as the stabilizer code given by the $4t\times 2t$ matrix 
\begin{equation}\label{Eq: aniso-ext-stab-mat}
    \A_\mathrm{ext} = \begin{pmatrix}
        \A^Z_{t\times t} & 0 \\
        (1-z)\mathbf{1}_t & 0\\
        \hline
        0 & (1-\bar z) \mathbf{1}_t\\
        0 & -(\A^Z)^\dagger_{t\times t}
    \end{pmatrix}.
\end{equation}
Here, $z$ is the coordinate along the extended/anisotropic direction and the polynomial $1-z$ corresponds to Ising-type interactions between adjacent classical codes. It is easy to verify that $\A_\mathrm{ext}$ satisfies \eqref{Eq: comm-rel}

For example, the 2D toric code is an anisotropic extension of the 1D Ising model along the $y$ direction. This is easily seen by comparing \eqref{Eq: toric_code} with the anisotropic extension of \eqref{Ising-stabmat} and interpreting $y$ as the anisotropic direction.

The GSDs of a classical code $\A$ and its anisotropic extension $\A_\mathrm{ext}$ are related as
\begin{equation}\label{Eq: GSD-aniso-class}
    \mathrm{GSD}_{\A_\mathrm{ext}}=\mathrm{GSD}_{\A}^2.
\end{equation} This follows from a doubling of the number of logical operators from the classical code to its anisotropic extension. 
For example, the 1D Ising model on a periodic chain is two-fold degenerate while the GSD of the 2D toric code on a periodic square lattice is $4$, which is indeed $2^2$.

As the structure of the stabilizer matrix \eqref{Eq: aniso-ext-stab-mat} suggests, the anisotropic extension of a classical code enjoys a self-duality that exchanges the $Z$-type and $X$-type stabilizer terms. In the example of the 2D toric code, this is the well-known $e\text{-}m$ duality that exchanges the $e$ and $m$ anyons.

\section{Anisotropic $\Z_N$ Laplacian model on regular lattices\label{Sec: anisotropic}}
The anisotropic $\Z_N$ Laplacian model \cite{PhysRevB.107.125121,10.21468/SciPostPhys.14.5.106} is a stabilizer code  defined on the lattice $\Gamma\times C_{L_z}$ where $\Gamma$ is a simple, connected, undirected graph, and $C_{L_z}$ is a cycle graph (a circle along the $z$-direction) with the number of lattice sites in the $z$-direction being $L_z$. We use $(\vec i,z)$ to label the sites of the lattice and $(\vec i,z+\tfrac12)$ to label the $z$-links. We also use $\langle \vec i,\vec j\rangle$ to denote an edge of the graph $\Gamma$.

There are two sets of qudits: one on the sites of the  lattice, the other on the $z$-links connecting graphs with adjacent $z$-coordinates. The Hamiltonian is given by
\begin{equation}
    H=-\gamma_1\sum_{\vec{i},z}G(\vec{i},z)-\gamma_2\sum_{\vec{i},z}F\left(\vec{i},z+\tfrac{1}{2}\right)+\mathrm{h.c.},
    \label{Eq: anisotropic laplacian stabilizer}
\end{equation}
where
\begin{equation}
\begin{aligned}
    & G(\vec{i},z)=X_{\mathsf z,\left(\vec{i},z+\frac{1}{2}\right)}^\dagger X_{\mathsf z,\left(\vec{i},z-\frac{1}{2}\right)} \prod_{\vec{j}:\braket{\vec{i},\vec{j}}\in \Gamma}X_{(\vec{i},z)} X_{(\vec{j},z)}^\dagger,\\
    &F\left(\vec{i},z+\tfrac{1}{2}\right)=Z_{(\vec{i},z+1)}^\dagger Z_{(\vec{i},z)} \prod_{\vec{j}:\braket{\vec{i},\vec{j}}\in \Gamma} Z_{\mathsf z,\left(\vec{i},z+\frac{1}{2}\right)} Z_{\mathsf z,\left(\vec{j},z+\frac{1}{2}\right)}^\dagger.
     \label{Eq: stabilizers}
\end{aligned}
\end{equation}
Here $Z_{(\vec{i},z)},X_{(\vec i,z)}$ are the $\Z_N$  clock and shift operators acting on the qudit on the site $(\vec{i},z)$, whereas $Z_{\mathsf z,\left(\vec{i},z+\frac{1}{2}\right)},X_{\mathsf z,\left(\vec{i},z+\frac{1}{2}\right)}$ act on the qudit on the $z$-link $\left(\vec{i},z+\frac{1}{2}\right)$. 

The ground states of the anisotropic Laplacian model satisfy $G=F=1$. To count the number of ground states, we need to count the number of independent logical operators. The logical operators take the following form\footnote{It is straightforward to check that they commute with the stabilizers of anisotropic Laplacian models}:
\begin{equation}
\begin{aligned}
    & W_z(\vec{i})=\prod_z Z_{\mathsf z,\left(\vec{i},z+\frac{1}{2}\right)},\quad \Tilde{W}_z(\vec{i})=\prod_z X_{\left(\vec{i},z\right)}, \\
    & W(h;z)=\prod_{\vec{i}} Z_{\left(\vec{i},z\right)}^{h(\vec{i})},\quad \Tilde{W}(h;z+\tfrac12)=\prod_{\vec{i}} X_{\mathsf z,\left(\vec{i},z+\frac12\right)}^{h(\vec{i})},
    \label{Eq: anisotropic logicals}
\end{aligned}
\end{equation}
where $h(\vec{i})$ is a $\Z_N$-valued discrete harmonic function on the graph, i.e., \begin{equation}\label{Footnote: Delta_L}
    \Delta_L h(\vec{i})=\sum_{\vec j:\braket{\vec i,\vec j}\in \Gamma}[h(\vec i)-h(\vec j)] = 0\mod N.
\end{equation}
Here, $\Delta_L$ is the discrete Laplacian operator. By counting the number of independent Heisenberg algebras formed by these logical operators, it was shown in \cite{PhysRevB.107.125121} that the GSD depends only on the graph $\Gamma$ and is given by
\begin{equation}
    \mathrm{GSD}=\prod_{a=1}^{\mathsf N}\gcd(N,r_a)^2=|\mathrm{Jac}(\Gamma,N)|^2,
\end{equation}
where $r_a$'s are the invariant factors in the Smith normal form of the Laplacian matrix of $\Gamma$, $|\mathrm{Jac}(\Gamma,N)|$ is the order of the mod-$N$ reduction of the Jacobian group of $\Gamma$, and $\mathsf N$ is the number of vertices of $\Gamma$. The square comes from the fact that we have two sets of Heisenberg algebras $\{W_z, \tilde{W}(h)\}$ and $\{W(h), \tilde{W}_z\}$. 

Note that the anisotropic Laplacian model is the anisotropic extension of the Laplacian model along the $z$ direction. The Laplacian model is a classical code defined on the graph $\Gamma$ by the Hamiltonian
\begin{equation}
    H=-\gamma\sum_{\vec i}\prod_{\vec{j}:\braket{\vec{i},\vec{j}}\in \Gamma}Z_{\vec{i}} Z_{\vec{j}}^\dagger+\mathrm{h.c.},
    \label{Eq: laplacian stabilizer}
\end{equation}
where $Z_{\vec i}$ is the $\Z_N$ clock operator acting on the qudit on the vertex $\vec i$ of the graph $\Gamma$.

The anisotropic Laplacian model has point-like excitations, which are violations of the stabilizers $G$ and $F$. Due to the self-duality that exchanges the $G$ and $F$ terms, these two kinds of excitations have the same mobility. In particular, they are both $z$-lineons because they can move in the $z$-direction via open versions of the string operators $W_z$ and $\tilde W_z$, respectively. However, their mobility along the graph is nontrivial and is closely related to the mobility of the point-like excitations in the Laplacian model.

\subsection{On the extended triangular lattice \label{subsection: anisotropic laplacian triangular}}
\begin{figure}
    \centering
    \includegraphics[scale=1]{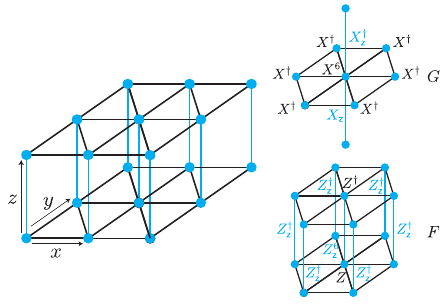}
    \caption{The prismatic lattice $\Gamma\times C_{L_z}$, 
    when the base graph $\Gamma$ is the 2D triangular lattice, with periodic boundary conditions. $C_{L_z}$ is shown as the blue links along $z$ direction. The two stabilizer generators are shown as $G$ and $F$. The blue variables are on the blue links ($z$-links) while the black ones are on the blue vertices (sites).}
    \label{fig: triangular lattice}
\end{figure}

In this section, we consider the anisotropic $\Z_N$ Laplacian model on the extended triangular lattice, referred to as the triangular-based anisotropic Laplacian model. Geometry of the 3D lattice is shown in Fig. \ref{fig: triangular lattice}. Periodic boundary conditions are implemented such that $(x,y,z)\sim (x+L_x,y,z)\sim (x,y+L_y,z)\sim (x,y,z+L_z)$. And the stabilizer terms \eqref{Eq: stabilizers} are given by
\begin{equation}
\begin{aligned}
    & G_\mathrm{tri}(\vec{i},z) = X_{\mathsf z,\left(\vec{i},z+\frac{1}{2}\right)}^\dagger X_{\mathsf z,\left(\vec{i},z-\frac{1}{2}\right)} X_{(\vec{i},z)}^6\prod_{\vec{j}\in \hexagon}X_{(\vec{j},z)}^\dagger, \\
    & F_\mathrm{tri}\left(\vec{i},z+\tfrac{1}{2}\right) = Z_{\left(\vec{i},z+1\right)}^\dagger Z_{\left(\vec{i},z\right)} Z_{\mathsf z,\left(\vec{i},z+\frac{1}{2} \right)}^6 \prod_{\vec{j}\in \hexagon} Z_{\mathsf z,\left(\vec{j},z+\frac{1}{2} \right)}^\dagger,
    \label{Eq: 3D anisotropic Laplacian triangular}
\end{aligned}
\end{equation}
where the product is taken over the six nearest-neighbouring sites around the site $\vec{i}$. 
The stabilizers can be rewritten in terms of polynomial matrix\footnote{From \eqref{Eq: anisotropic laplacian stabilizer}, we notice that there are four terms meaning that the stabilizer matrix should be $4\times 4$. However we can apply the column operations introduced in Sec. \ref{subsection: element_transf} to eliminate one of $F$ and $F^\dagger$, as well as one of $G$ and $G^\dagger$. So the matrix is reduced to $4\times2$. In the following, we will always use this elimination.},
\begin{widetext}
    \begin{equation}
        \A_\mathrm{tri} = \begin{pmatrix}
            6-(x+y+\Bar{x}+\Bar{y}+x\Bar{y}+\Bar{x}y)  & 0 \\
            1-z & 0 \\
            \hline
            0 & 1-\Bar{z} \\
            0 & -6+(x+y+\Bar{x}+\Bar{y}+x\Bar{y}+\Bar{x}y) \\
        \end{pmatrix}.
        \label{Eq: tri_matrix}
    \end{equation}
\end{widetext}

When $N=2$, in addition to the string-like logical operators $W_z$ and $\tilde W_z$ in the $z$ direction, there are string-like logical operators in the $xy$ plane:
\begin{equation}\label{Eq: logical_xy_tri}
    \tilde W_C(z+\tfrac12) = \prod_{(n,m)\in C} X_{\mathsf z,(n,m,z+\frac12)},
\end{equation}
where $C$ is a (non-contractible) rigid straight line through the sites in the $(0,1)$, $(1,0)$, or $(1,-1)$ direction. The number of such lines is $L_x$, $L_y$, and $\gcd(L_x,L_y)$, respectively. Similarly, there are string-like logical operators $W_C$ for each such curve. Moreover, the logical operators $\tilde W_C$ are independent of each other except for the two relations
\begin{equation}
    \prod_{C_y} \tilde W_{C_y} = \prod_{C_x} \tilde W_{C_x} = \prod_{C_{x\bar y}} \tilde W_{C_{x\bar y}} = \prod_{n,m} X_{\mathsf z,(n,m,z+\frac12)},
\end{equation}
where $C_y$, $C_x$, and $C_{x\bar y}$ are the straight lines in the $(0,1)$, $(1,0)$, and $(1,-1)$ directions, respectively. Together with the $W_z$ and $\tilde W_z$ operators, they lead to a GSD given by
\begin{equation}\label{Eq: log2 GSD tri}
    \log_2 \mathrm{GSD_{tri}} = 2[L_x + L_y +\gcd(L_x,L_y) - 2].
\end{equation}

When $N>2$, there is no simple structure to the logical operators in the $xy$ plane. Instead, we calculate the ground-state degeneracy using the techniques from commutative algebra.

\begin{figure}
    \includegraphics[scale=0.4]{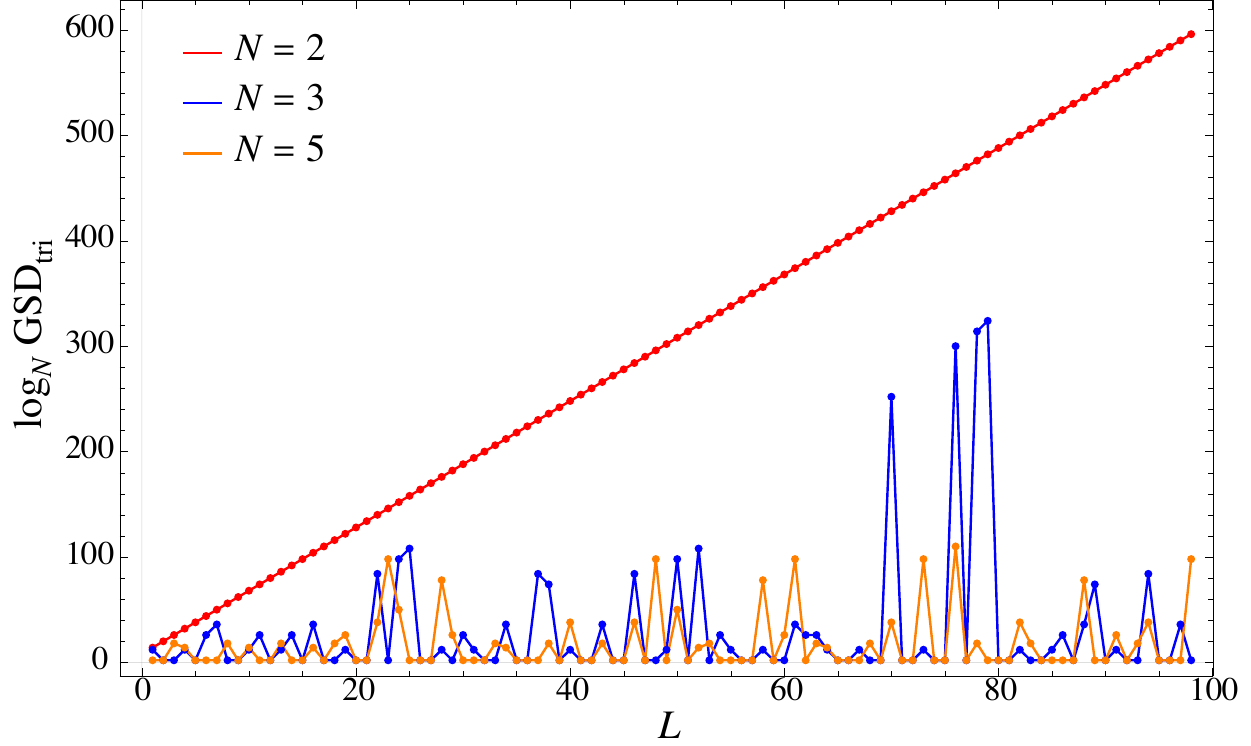}
    \caption{The logarithm of the ground-state degeneracy of the triangular-based anisotropic $\Z_N$ Laplacian model for $N=2$ (red), $3$ (blue) and $5$ (orange). The system size is taken to be $L\times L\times L_z$ with $3\leq L\leq 100$. When $N=2$, $\log_2\mathrm{GSD}$ grows linearly in $L$. However, when $N=3,5$, we see that  $\log_N \mathrm{GSD}$ fluctuates sharply with $L$. }
    \label{fig:logGSD_Tri}
\end{figure}

In particular, we show that when $N=p$ is an odd prime, for $L_x=p^{k_x}q^m$ and $L_y=p^{k_y}q^m$, where  $k_x,k_y,m\geq 0$ are integers and $q\neq p$ is another odd prime such that $p$ is a primitive root modulo $q^m$ for $m\geq 1$, we have
\begin{equation}
    \log_p\mathrm{GSD}_\mathrm{tri}=
    \begin{cases}
        4\times 3^{\min(k_x,k_y)}, & p=3,\\
        2[2p^{\min(k_x,k_y)}-\delta_{k_x,k_y}],  & p>3.
    \end{cases}
    \label{Eq: logpGSD_tri}
\end{equation}
See Appendix \ref{appendix: gsd_ca} for details\footnote{Throughout Appendix \ref{appendix: gsd_ca}, we calculate the GSD of the (classical) Laplacian model. The GSD of the anisotropic Laplacian model follows from the relation \eqref{Eq: GSD-aniso-class}.}. The numerical calculation of GSD is shown in Fig. \ref{fig:logGSD_Tri}. The dramatic fluctuation of GSD with the system size when $p>2$ suggests type-II fracton order, which we confirm by studying the mobility of excitations below.

\begin{figure}
    \centering
    \includegraphics[scale=0.7]{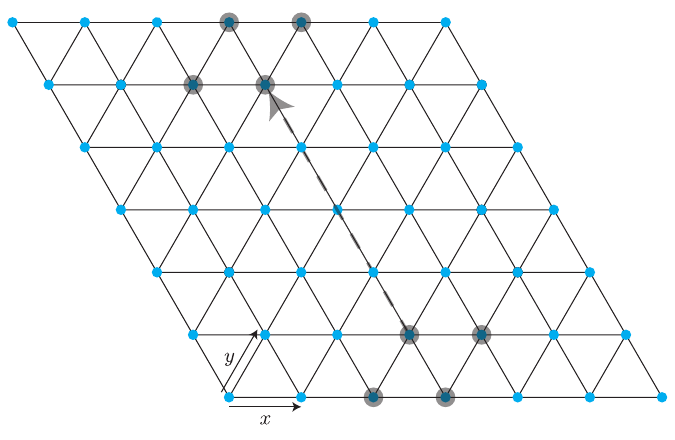}
    \caption{The mobility of quadrupole of $z$-lineons (gray dots) in the triangular-based anisotropic $\Z_2$ Laplacian model. The gray dashed arrow in the $(-1,1)$ direction indicates the string operator that moves the quadrupole located at relative positions $(0,0)$, $(1,0)$, $(0,1)$, and $(1,1)$. This string operator is an open version of the logical operator $W_C$ or $\tilde W_C$.}
    \label{fig: tri_mobility}
\end{figure}

As explained before, the triangular-based anisotropic Laplacian model has string-like logical operators along the $z$-directions, and therefore hosts $z$-lineons. The mobility of $z$-lineons on the 2D base lattice depends on $N$. For $N=2$, while a single $z$-lineon is immobile, a quadrupole of $z$-lineons can move along a line as depicted in Fig. \ref{fig: tri_mobility}.  This is related to the existence of the string-like logical operators $W_C$ and $\tilde W_C$ in the $xy$ plane, whose open versions ``move'' a quadrupole of $z$-lineons that violate $G$ and $F$ terms, respectively. On the other hand, for $N=p$ an odd prime, any finite set of $z$-lineons cannot move, unless they are created locally. We prove these statements in Appendix \ref{appendix: mobility_tri_laplacian}.

Lastly, we examine the robustness of the triangular-based anisotropic Laplacian model following the discussion in \cite{PhysRevB.107.125121}. A model is robust if there are no local deformations of the Hamiltonian that commute with the Hamiltonian and lift the ground state degeneracy, i.e., if there are no logical operators with local (finite) support in the thermodynamic (infinite volume) limit.

Recall that the logical operators take the form \eqref{Eq: anisotropic logicals}. Obviously, $W_z(\vec{i})$ and $\Tilde{W}_z(\vec{i})$ are supported over $L_z$ so they have infinite support in this limit. To investigate the behaviour of $W(h)$ and $\Tilde{W}(h)$ at a fixed $z$, we turn to look at the discrete harmonic function $h(\vec i) = h(x,y)$ because the support of these logical operators is the same as the support of $h(x,y)$. On the infinite triangular lattice, there is no nontrivial finitely supported discrete harmonic function. We prove this fact in Appendix \ref{appendix: robust-kagome}. Therefore, $W(h)$ and $\tilde W(h)$ do not have local support, and hence the triangular-based anisotropic Laplacian model is robust.

\subsection{On the extended honeycomb lattice \label{subsection: anisotrpoic laplacian hc}}
\begin{figure}
    \centering
    \includegraphics[scale=1.55]{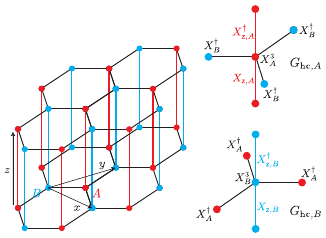}
    \caption{The prismatic lattice $\Gamma \times C_{L_z}$ whose base graph $\Gamma$ is the 2D honeycomb lattice with periodic boundary conditions. The two sublattices $A$ and $B$ are shown as the red and blue dots, and the $z$-links are colored accordingly. For each sublattice $i$, there are two conjugate stabilizers: $G_{\mathrm{hc},i}$ and $F_{\mathrm{hc},i}$ (we show only the former). As before, the colored variables are on the $z$-links, whereas the black variables are on the sites.}
    \label{fig: honeycomb lattice}
\end{figure}
Now we turn to discuss the anisotropic Laplacian model based on the honeycomb lattice. Again, we take the prismatic lattice with periodic boundary conditions shown in Fig. \ref{fig: honeycomb lattice}. Notice that there are two sublattices $A$ and $B$ so we have two different types of stabilizers $F_{A,B}$ and $G_{A,B}$. We denote the $\Z_N$ variables living on the sublattice $A$ as $X_{\mathsf z,A}$ ($Z_{\mathsf z,A}$) and $X_{A}$ ($Z_{A}$) and similarly for the other sublattice. Then the stabilizers on the $A$ sublattice are given as
\begin{equation}
\begin{aligned}
    & G_{\mathrm{hc},A}(\vec{i},z)=X_{\mathsf z,A,\left(\vec{i},z+\frac{1}{2}\right)}^\dagger X_{\mathsf z,A,\left(\vec{i},z-\frac{1}{2}\right)} X_{A,(\vec{i},z)}^3\prod_{\vec{j}\in \triangle}X_{B,(\vec{j},z)}^\dagger, \\
    & F_{\mathrm{hc},A}\left(\vec{i},z+\tfrac{1}{2}\right)= Z_{A,(\vec{i},z+1)}^\dagger Z_{A,(\vec{i},z)} Z_{\mathsf z,A,\left(\vec{i},z+\frac{1}{2}\right)}^3 \\
    &\quad \quad \quad \quad \quad \quad \quad \quad \;\times\prod_{\vec{j}\in \triangle}Z_{\mathsf z,B,\left(\vec{j},z+\frac{1}{2}\right)}^\dagger,
\end{aligned}
\end{equation}
where the product is taken over the three nearest-neighbouring sites $\vec{j}$ around the site $\vec{i}$, and similarly for $G_{\mathrm{hc},B}$ and $F_{\mathrm{hc},B}$. In the polynomial formalism, the Hamiltonian is associated with the matrix
\begin{equation}
    \A_\mathrm{hc}=\begin{pmatrix}
        3 & -1-\Bar{x}-\Bar{y} & 0 & 0 \\
        -1-x-y & 3 & 0 & 0 \\
        1-z & 0 & 0 & 0 \\
        0 & 1-z & 0 & 0 \\
        \hline
        0 & 0 & 1-\bar{z} & 0 \\
        0 & 0 & 0 & 1-\bar{z} \\
        0 & 0 & -3 & 1+\Bar{x}+\Bar{y} \\
        0 & 0 & 1+x+y & -3
    \end{pmatrix}.
\end{equation}

When $N$ is coprime to $3$, the honeycomb anisotropic $\Z_{N}$ Laplacian model is equivalent to that on the triangular lattice, and hence should share the same properties. The explicit sequential local unitary transformation that relates these two models is given by
\begin{equation}
    \begin{aligned}
        &\mathrm{Col}(2,1,k(1+\bar{x}+\bar{y}))~\mathrm{Col}(4,3,k(1+\bar{x}+\bar{y}))
        \\
        &\times\mathrm{CNOT}(3,4,-k(1+\bar{x}+\bar{y}))
        \\
        &\times \mathrm{CNOT}(3,1,-k(1-z))
        \\
        &\times \mathrm{CNOT}(2,1,k(1+x+y)),
    \end{aligned}
\end{equation}
where $k$ is an integer given by $3k = 1\mod N$. We further multiply the first and seventh rows by $k$, and second and eighth rows by $3$, which is complemented by multiplying the fifth and third rows by $3$, and sixth and fourth rows by $k$. Ignoring two decoupled sets of qudits leaves two blocks which, after a rearrangement, form the stabilizer matrix of the anisotropic triangular-based Laplacian model \eqref{Eq: tri_matrix}.

For $N=3$, the monomial entry 3 is reduced to 0, and the honeycomb-based anisotropic Laplacian model is decoupled into two copies of the 3+1D  anisotropic $\Z_3$ fractal spin liquid (FSL) with the stabilizer matrix
\begin{equation}\label{Eq: fsl-stab-mat}
    \A_\mathrm{FSL}=\begin{pmatrix}
        1+x+y & 0 \\
        1-z & 0 \\
        \hline
        0 & 1-\Bar{z} \\
        0 & -1-\Bar{x}-\Bar{y} \\
    \end{pmatrix},
\end{equation}
which is the anisotropic extension of the 2D fractal spin liquid model with stabilizer matrix $(1+x+y,0)^{\mathsf T}$.

\begin{figure}
    \centering
    \includegraphics[scale=0.41]{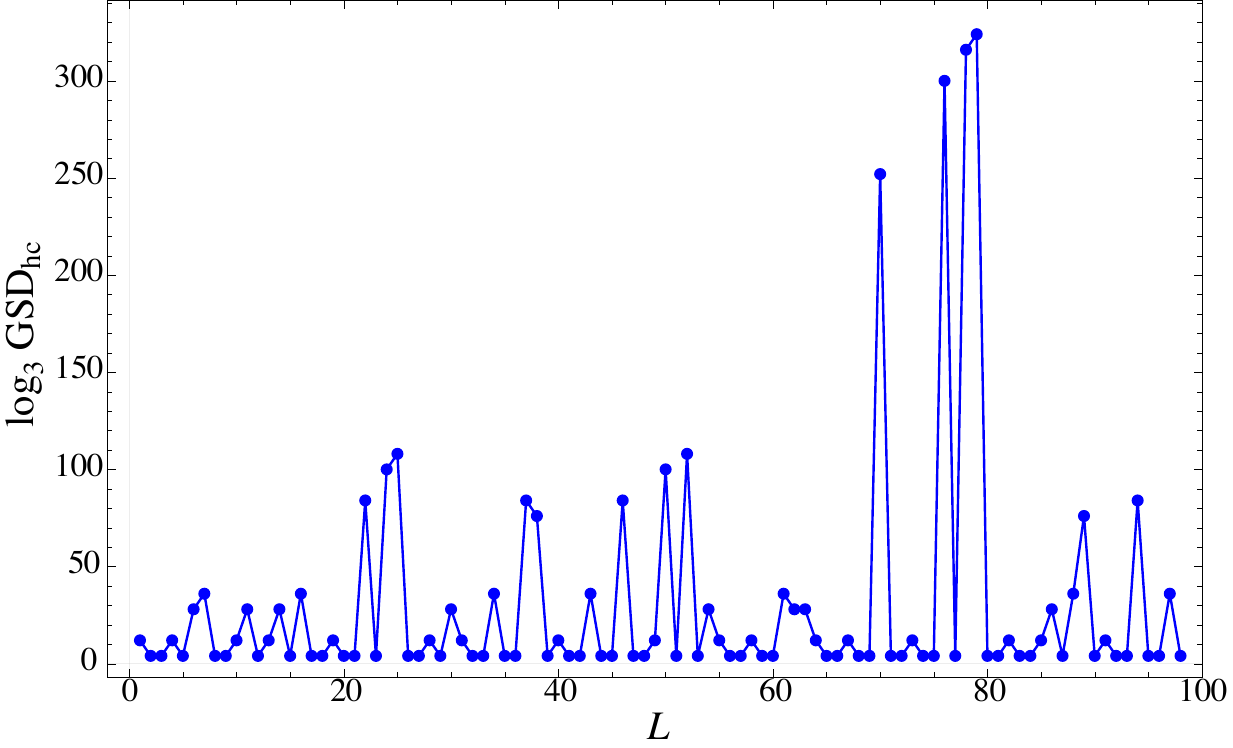}
    \caption{The logarithm of the ground-state degeneracy of the honeycomb-based anisotropic $\Z_3$ Laplacian model on the $L\times L\times L_z$ lattice with $3\leq L\leq 100$. It is clear that there is a dramatic fluctuation of GSD with increasing $L$.}
    \label{fig:logGSD_hc}
\end{figure}

In Appendix \ref{appendix: gsd_ca}, we calculate the GSD of the anisotropic $\Z_3$ fractal spin liquid as
\begin{equation}
    \log_3\mathrm{GSD}_\mathrm{FSL}=2\times 3^{\min(k_x,k_y)},
\end{equation}where $k_x$ and $k_y$ are defined the same as in \eqref{Eq: logpGSD_tri}. The GSD of $\Z_3$ honeycomb anisotropic Laplacian is thus given by
\begin{equation}
    \log_3\mathrm{GSD}_\mathrm{hc}=4\times 3^{\min(k_x,k_y)}.
    \label{Eq: log3GSD hc}
\end{equation}
Curiously, this GSD matches with the GSD of the triangular-based anisotropic $\Z_3$ Laplacian model for these special values of $L_x$ and $L_y$. However, numerical computation shows that they do not always match for other values of $L_x$ and $L_y$. Therefore, the two models are not equivalent for $N=3$. The numerical plot of the $\mathrm{GSD_{hc}}$ is shown in Fig. \ref{fig:logGSD_hc}.

\begin{figure}
    \centering
    \includegraphics[scale=1.15]{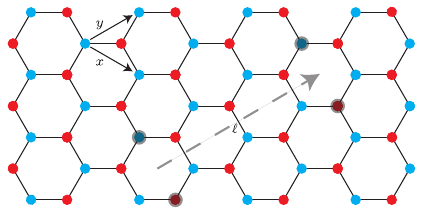}
    \caption{The mobility of a dipole of $z$-lineons (gray dots) in the honeycomb-based anisotropic $\Z_2$ Laplacian model. The dashed line represents the open version of the string-like logical operator $W_C$ or $\tilde W_C$. }
    \label{fig: honeycomb_mobility}
\end{figure}

The mobility of $z$-lineons in this model is similar to that of the triangular-based anisotropic Laplacian model when $N$ is coprime to $3$. For example, when $N=2$, a dipole of $z$-lineons can move in the direction orthogonal to its orientation in the $xy$ plane as depicted in Fig. \ref{fig: honeycomb_mobility}. Relatedly, similar to \eqref{Eq: logical_xy_tri}, there are string-like logical operators in the $xy$ plane:
\begin{equation}
    \tilde W_C(z + \tfrac{1}{2}) = \prod_{\ell\in C}X_{\mathsf z,A,\left(\ell, z+\frac{1}{2}\right)} X_{\mathsf z,B,\left(\ell, z+\frac{1}{2}\right)},
\end{equation}
where $C$ is a rigid straight line through the centers of the hexagons in the $(0,1)$, $(1,0)$, or $(1,-1)$ direction, the product is over all links $\ell$ pierced by $C$, and $X_{\mathsf z,A,\left(\ell, z+\frac{1}{2}\right)}$ and $ X_{\mathsf z,B,\left(\ell, z+\frac{1}{2}\right)}$ act on the two qubits at the ends of the link $\ell$. Similarly, there are string-like logical operators $W_C$. The open versions of $W_C$ and $\tilde W_C$ ``move'' the dipoles of $z$-lineons that violate $G$ and $F$ terms, respectively, as in Fig. \ref{fig: honeycomb_mobility}.

For $N=3$, by the same argument as in Appendix \ref{appendix: mobility_tri_laplacian}, it is straightforward to show that any finite set of $z$-lineons is immobile in the $xy$ plane, except when they are created locally. This result follows from the fact that $x^{n_0}y^{m_0}-1$ is not a multiple of $f(x,y)=1+x+y\mod 3$.

The robustness of the anisotropic Laplacian model on honeycomb lattice follows from the same argument as that on the triangular lattice, which is spelled out in Appendix \ref{appendix: robust-kagome}.

\subsection{On the extended Kagome lattice \label{subsection: anisotropic Laplacian_Kagome}}
\begin{figure*}[]
    \includegraphics[scale=0.63]{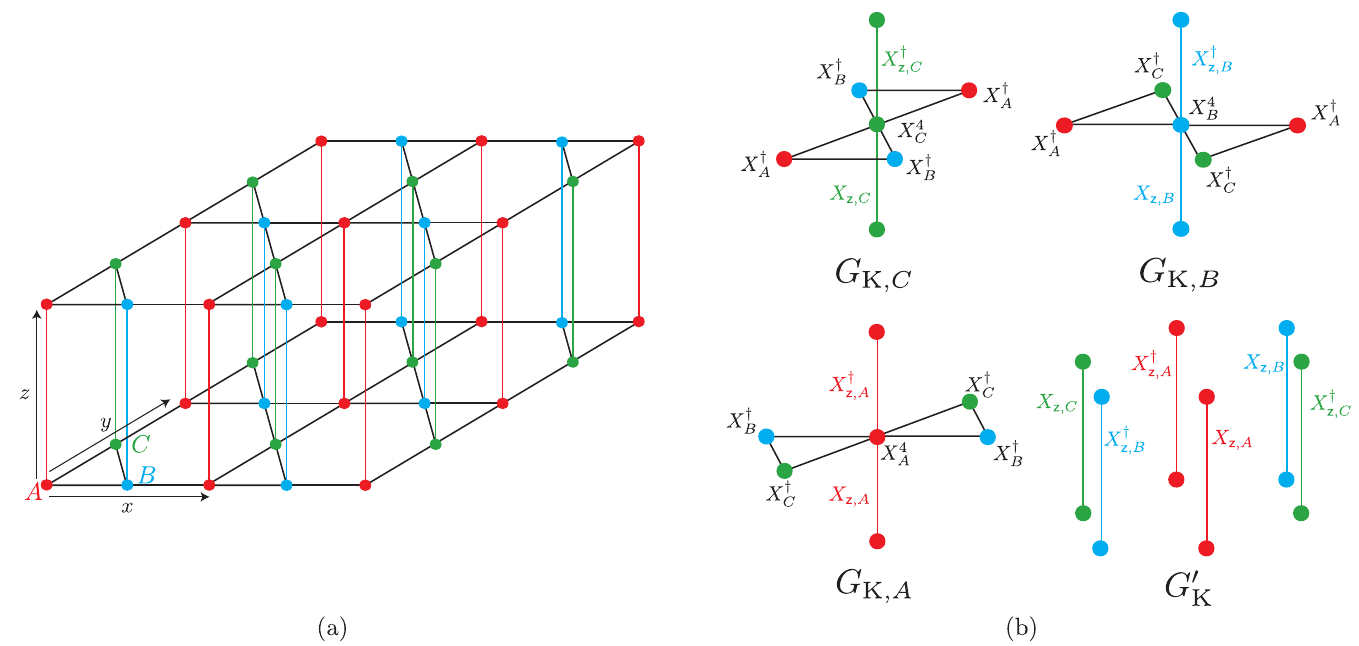}
    \caption{(a) The prismatic lattice $\Gamma \times C_{L_z}$ with Kagome-lattice base with periodic boundary conditions. There are three sublattices denoted as red ($A$), blue ($B$) and green ($C$) dots respectively. (b) $G_{\mathrm{K},A}$, $G_{\mathrm{K},B}$, and $G_{\mathrm{K},C}$ are the stabilizer generators of the Kagome-based anisotropic Laplacian model (we do not show the $F$ terms). When $N$ is a multiple of $2$ or $3$, this model is not robust. Relatedly, there is an additional local term $G_\mathrm{K}'$ (shown for $N=2,3$) that acts nontrivially on the ground-state subspace and commutes with all the terms of the Hamiltonian.}
    \label{fig:Kagome}
\end{figure*}

We now turn to the anisotropic Laplacian model based on the Kagome lattice, which contains three sublattices $A$, $B$ and $C$ as shown in Fig. \ref{fig:Kagome}. Using the similar conventions in Sec. \ref{subsection: anisotrpoic laplacian hc}, we denote $\Z_N$ operators as $X_{\mathsf z,i}$ ($Z_{\mathsf z,i}$) and $X_{i}$ ($Z_{i}$) where $i=A,B,C$. The stabilizers on the sublattice $A$ can be written as
\begin{equation}
\begin{aligned}
    & G_{\mathrm{K},A}(\vec{i},z)=X_{\mathsf z,A,\left(\vec{i},z+\frac{1}{2}\right)}^\dagger X_{\mathsf z,A,\left(\vec{i},z-\frac{1}{2}\right)} X_{A,\left(\vec{i},z\right)}^4  \\
    &\quad \quad \quad \quad \quad \;\; \times\prod_{\vec{j} \in \square} X^\dagger_{B/C,(\vec{j},z)},  \\
    & F_{\mathrm{K},A}\left(\vec{i},z+\tfrac{1}{2}\right)= Z_{A,(\vec{i},z+1)}^\dagger Z_{A,(\vec{i},z)} Z_{\mathsf z,A,\left(\vec{i},z+\frac{1}{2}\right)}^4 \\
    &\quad \quad \quad \quad \quad \quad \quad \;\;\;\;\times\prod_{\vec{j}\in \square}Z_{\mathsf z,B/C,\left(\vec{j},z+\frac{1}{2}\right)}^\dagger,
\end{aligned}
\end{equation}
where the product is taken over the four nearest-neighbouring sites around the site $\vec{i}$ and similarly for the stabilizers defined on the other two sublattices.

\begin{figure}
    \centering
    \includegraphics[scale=0.415]{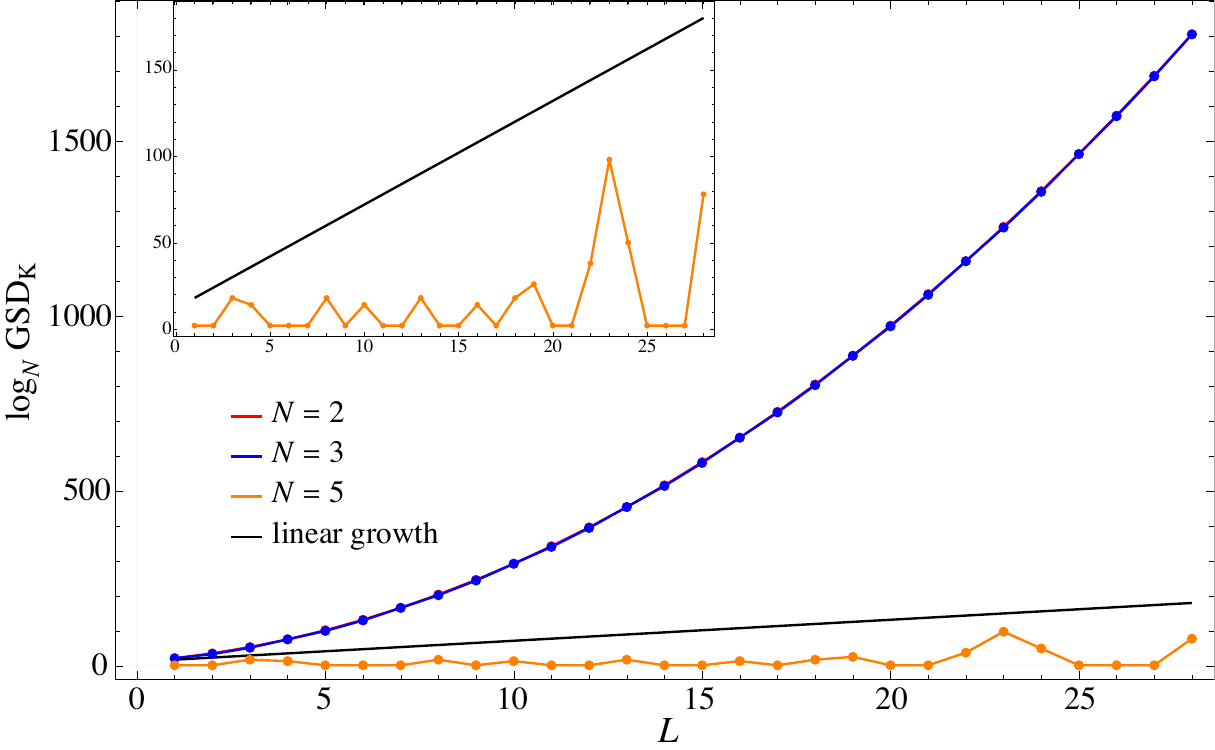}
    \caption{The logarithm of the ground-state degeneracy of the Kagome-based anisotropic Laplacian model on the $L\times L\times L_z$ lattice with $3\leq L\leq 30$ for $N=2$ (red), $3$ (blue), and $5$ (orange). The blue line is almost on top of the red one, so the latter is not visible. The black line shows a linear function of $L$. The $\Z_{2,3}$ model is not robust as indicated by the faster-than-linear growth of $\log_{2,3} \mathrm{GSD}$ with $L$. The $\Z_5$ model is robust as suggested by the at-most-linear behaviour of $\log_5\mathrm{GSD}$ with the system size.}
    \label{fig:logGSD_Kagome}
\end{figure}

Contrary to what was discussed in the Footnote 15 of \cite{PhysRevB.107.125121}, the anisotropic Laplacian model based on the Kagome lattice is not robust for all $N$. A powerful test of robustness is the growth of the ground-state degeneracy with system size. It is shown in \cite{10.21468/SciPostPhys.10.1.011} that the logarithm of the ground-state degeneracy for systems with homogeneous topological order on an arbitrary closed Riemannian manifold of spatial dimension $D$ cannot grow faster than $L^{D-2}$, where $L$ is the linear size of the system. When $D=3$, $\log_N\mathrm{GSD}$ is restricted from growing faster than linearly in $L$. We find that the Kagome anisotropic $\Z_{2,3}$ Laplacian model violates this bound (see the red and blue plots in Fig. \ref{fig:logGSD_Kagome}). Upon examining the logical operators more closely, we find that there exist local operators that commute with all terms of the Hamiltonian in the $\Z_{2,3}$ model. They are shown as $G_\mathrm{K}'$ in the lower-right panel in Fig. \ref{fig:Kagome}(b). Similarly, there are operators $F_\mathrm{K}'$ that also commute with all terms of the Hamiltonian. More generally, $(G'_\mathrm{K})^k$ and $(F_\mathrm{K}')^k$ are local logical operators whenever $N=2k$ or $N=3k$, i.e., whenever $N$ is a multiple of $2$ or $3$.

On the other hand, looking at $N=5$ in the inset of Fig. \ref{fig:logGSD_Kagome}, we see that $\log_N\mathrm{GSD}$ is bounded by a linear function of $L$, suggesting that the $\Z_5$ model is robust. Indeed, we prove in Appendix \ref{appendix: robust-kagome} that the $\Z_N$ model has no local logical operators when $N$ is coprime to $6$.

It follows from the last two paragraphs that the Kagome-based anisotropic $\Z_N$ Laplacian model is robust if and only if $N$ is coprime to $6$.

\section{Entanglement renormalization of the anisotropic Laplacian models\label{Sec: ER_Laplacian}}
In this section, we study the entanglement renormalization of the anisotropic $\Z_2$ and $\Z_3$ Laplacian models on the extended square, triangular and honeycomb lattices respectively.
Anisotropic models have rigid string operators along the $z$-direction, so they always host $z$-lineons. These include the fractal spin liquid \cite{PhysRevB.88.125122,doi:10.1080/14786435.2011.609152}, some members of the cubic code (CC) family \cite{PhysRevA.83.042330} ($\mathrm{CC}_{5,6,9}$, $\mathrm{CC}_{11-17}$) and the anisotropic Laplacian model \cite{PhysRevB.107.125121}. Thus, they are invariant under the ER along the $z$-direction \cite{PhysRevResearch.2.033021}. Therefore, in the following discussions, we will focus only on the ER of the Laplacian models living on the base 2D lattices. The ER of anisotropic Laplacian models can be obtained by anisotropic extension (described in Sec. \ref{subsection: anisotropic extension}) of all the models involved in the ER. All models are placed on an infinite lattice for the purpose of ER calculations. The explicit calculations can be found in the Mathematica file attached to this submission.

All models studied in this paper turn out to be bifurcating under an appropriate coarse-graining factor. We investigate the relationships between the GSDs before and after ER as a consistency check. We place the models on finite lattice with periodic boundary conditions to compute the GSDs. For convenience, in the following, we denote the original models as $\A$ codes and the associated child models after the ER as $\B$ codes.

Throughout this paper, we will use the notation
\begin{equation}
    H(a)\xrightarrow{c\times c} H_{1}(a_1)+H_2(a_2)+\cdots
\end{equation}
to denote the ER flow where the model $H$ with the lattice spacing $a$ flows into the models $H_1,H_2,\ldots$ with the lattice spacing $ca$ respectively, under the coarse-graining by $c$ along $x$ and $y$ directions. 

\subsection{Entanglement renormalization}
\label{subsection: ER general}

The basic idea of entanglement renormalization transformation is to disentangle the short-range degrees of freedom by applying local unitary transformations \cite{PhysRevLett.99.220405,PhysRevLett.100.070404,PhysRevResearch.2.033021,PhysRevB.89.075119}. The general procedure shown in \eqref{Eq: bifurecating_relation} consists of the following steps: (i) coarse-grain the lattice by enlarging the unit cell by the factor $c>1$, (ii) apply local unitary transformations defined in Sec. \ref{subsection: element_transf}, and (iii) factor out the disentangled trivial qudits.

Coarse-graining is an operation that groups sites together to form a larger unit cell \cite{PhysicsPhysiqueFizika.2.263}. In $D$ dimensions, coarse-graining in all directions by a factor of $c$ will enlarge the number of qudits per site and the number of stabilizer generators per site by a factor of $c^D$. In the polynomial language, under the coarse-graining by $c$ in $x$ direction while keeping the other directions the same, the polynomials in the stabilizer matrix transform as
\begin{equation}
    x\mapsto\begin{pmatrix}
        0 & x' \\
        \mathbf{1}_{c-1} & 0 \\
    \end{pmatrix},\quad
    y\mapsto y\mathbf{1}_c,\quad
    z\mapsto z\mathbf{1}_c,
\end{equation}
where $x'=x^c$ is the new variable in the $x$ direction.

Typically, type-II fracton models are bifurcating fixed points only for certain values of the coarse-graining factor $c$. This is in contrast to ordinary topologically-ordered models and type-I fracton models, which are bifurcating fixed points for any choice of $c$.

\subsection{On the square lattice\label{subsection: ER Sq}}
Let us first consider the $\Z_2$ Laplacian model defined on the 2D square lattice $\A_\mathrm{sq}^{\Z_2}$. It is equivalent to the $\Z_2$ plaquette Ising model \cite{PhysRevB.103.195113,10.21468/SciPostPhys.10.2.027} on the tilted lattice as shown in Fig. \ref{fig: tilted} whose anisotropic extension is called the anisotropic lineon model  \cite{SHIRLEY2019167922,PhysRevB.103.205116}.  The ER properties of the anisotropic lineon model have been studied in \cite{SHIRLEY2019167922}; it was shown that the anisotropic lineon model flows to itself and a stack of toric codes if coarse grained along the $x$ or $y$ directions. 

Since $\A_\mathrm{sq}^{\Z_2}$ is mapped to $45^{\circ}$-tilted Ising plaquette model, coarse-graining of the former should be done along both $x$ and $y$ direction with the factor of 2. This can also be confirmed by calculating the charge annihilator and it is found that the annihilator is invariant under this coarse-graining. Concretely, the stabilizer matrix is given by
\begin{equation}
    \A^{\Z_2}_\mathrm{sq}=
    \begin{pmatrix}
        x+y+x^2y+xy^2 \\\hline 0
    \end{pmatrix}.
\end{equation}
The ``coarse-grained" (but without any additional local unitary transformation) version of it is given by
\begin{equation}
    \begin{pmatrix}
        y+xy & 1+y & 0 & 0 \\
        x+xy & 1+x & 0 & 0 \\
        0 & 0 & y+xy & 1+y \\
        0 & 0 & x+xy & 1+x \\
        \hline\multicolumn{4}{c}{0}
    \end{pmatrix}.
\end{equation}
We have not done any nontrivial operation yet---the coarse-graining step is just a redefinition of the unit cell---but we see that $\A_\mathrm{sq}^{\Z_2}$ is transformed into two copies of a ``new" model given by
\begin{equation}\label{Eq: Z2-sq-newmodel}
    \tilde{\A}^{\Z_2}_\mathrm{sq} = \begin{pmatrix}
        y+xy & 1+y \\
        x+xy & 1+x \\
        \hline\multicolumn{2}{c}{0}
    \end{pmatrix}.
\end{equation}
This is nothing but the $\Z_2$ plaquette Ising model on $45^\circ$-tilted square lattice. This transformation can be denoted as
\begin{equation}
    \A^{\Z_2}_\mathrm{sq}(a)= 2 \times \Tilde{\A}^{\Z_2}_\mathrm{sq}(2a),
    \label{Eq: A_sq=2A'_sq}
\end{equation}
where we use $=$ to imply no qubits are dropped. Eq. \eqref{Eq: A_sq=2A'_sq} simply reveals that there are two copies of the plaquette Ising model on the two sublattices shown in Fig. \ref{fig: tilted}.

Next, the ER transformation of $\Tilde{\A}^{\Z_2}_\mathrm{sq}$ is given by
\begin{equation}
    \Tilde{\A}^{\Z_2}_\mathrm{sq}(a)\xrightarrow{2\times 2} \Tilde{\A}^{\Z_2}_\mathrm{sq}(2a)+\mathrm{Is}^{\Z_2}_{1+xy}(2a)+\mathrm{Is}^{\Z_2}_{x+y}(2a),
    \label{Eq: ERG_A'_sq}
\end{equation}
where $\mathrm{Is}^{\Z_2}_{1+xy}=(1+xy,\;0)^\mathsf{T}$ is a stack of 1D Ising models extended along the $(1,1)$ direction and stacked along the $(1,-1)$ direction, and similar comments apply to $\mathrm{Is}^{\Z_2}_{x+y}=(x+y,\;0)^\mathsf{T}$. Combining \eqref{Eq: A_sq=2A'_sq} and \eqref{Eq: ERG_A'_sq}, we have
\begin{equation}
\begin{aligned}
    \A_\mathrm{sq}^{\Z_2}(a) & \xrightarrow{4\times 4} 2\Tilde{\A}_\mathrm{sq}^{\Z_2}(4a) + 2(\mathrm{Is}^{\Z_2}_{1+xy} + \mathrm{Is}^{\Z_2}_{x+y})(4a) \\
    & =\A_\mathrm{sq}^{\Z_2}(2a) + 2(\mathrm{Is}^{\Z_2}_{1+xy} + \mathrm{Is}^{\Z_2}_{x+y})(4a)
    \label{Eq: ERG_A_sq}
\end{aligned}
\end{equation}
Hence, the $\Z_2$ Laplacian model on the square lattice is a quotient bifurcating fixed point. Applying the anisotropic extension along the $z$ direction, the anisotropic $\Z_2$ Laplacian model bifurcates into itself and stacks of 2D toric codes.

The bifurcating results above can be verified by comparing the ground-state degeneracies. In the following, we use $k$ to denote $\log_N\mathrm{GSD}$. As reviewed in \eqref{Eq: logGSD_Z2_square}, the logarithm of GSD of the $\Z_2$ Laplacian model on the square lattice with $L_x=L_y=L$ is $2L$ for even $L$. Therefore, the GSDs satisfy 
\begin{equation}
    k_{\A,\mathrm{sq}}^{\Z_2}(4L) = k_{\A,\mathrm{sq}}^{\Z_2}(2L) + 2[k_{1+xy}^{\Z_2}(L) + k_{x+y}^{\Z_2}(L)].
\end{equation}
This is consistent with the ER flow \eqref{Eq: ERG_A_sq} because
\begin{equation}
    k_{1+xy}^{\Z_2}(L) = k_{x+y}^{\Z_2}(L) = L \times k_\mathrm{Is}^{\Z_2} = L.
\end{equation}

Now we consider the $\Z_3$ Laplacian model on the square lattice. The stabilizer matrix is given by
\begin{equation}
    \A_\mathrm{sq}^{\Z_3}=\begin{pmatrix}
        xy-x^2y-xy^2-x-y \\
        \hline 0 \\
    \end{pmatrix}.
\end{equation}
We show that the $\Z_3$ Laplacian model is also a quotient bifurcating fixed point under the coarse-graining by 3 with the ER flow
\begin{equation}\label{Eq: ER-sq-Z3-A}
    \A_\mathrm{sq}^{\Z_3}(a) \xrightarrow{3\times 3} \A_\mathrm{sq}^{\Z_3}(3a) + 2\B_\mathrm{sq}^{\Z_3}(3a),
\end{equation}
where
\begin{equation}
    \B_\mathrm{sq}^{\Z_3}=\begin{pmatrix}
        -x+y & 1-x \\
        -y+xy & 1-xy \\
        \hline
        \multicolumn{2}{c}{0}
    \end{pmatrix}.
    \label{Eq: B_square}
\end{equation}
Under the same coarse-graining factor,  the $\B$ code is a self-bifurcating fixed point, i.e.
\begin{equation}\label{Eq: ER-sq-Z3-B}
    \B_\mathrm{sq}^{\Z_3}(a) \xrightarrow{3\times 3} 3\B_\mathrm{sq}^{\Z_3}(3a).
\end{equation}

As a consistency check, we compute the ground-state degeneracy of the $\B$ code using the Gr\"obner basis technique for some special values of $L_x$ and $L_y$ (see Appendix \ref{appendix: B sq} for details). It is given by  
\begin{equation}
    k_{\B,\mathrm{sq}}^{\Z_3}=2\times 3^{\mathrm{min}(k_x,k_y)},
\end{equation}
where $k_x$ and $k_y$ are defined as in \eqref{Eq: logpGSD_tri}. For these special values of $L_x$ and $L_y$, it is easy to check that
\begin{equation}
     k_{\A,\mathrm{sq}}^{\Z_3}(3L_x,3L_y)=k_{\A,\mathrm{sq}}^{\Z_3}(L_x,L_y)+2 k_{\B,\mathrm{sq}}^{\Z_3}(L_x,L_y),
\end{equation}
and
\begin{equation}
    k_{\B,\mathrm{sq}}^{\Z_3}(3L_x,3L_y)=3 k_{\B,\mathrm{sq}}^{\Z_3}(L_x,L_y),
\end{equation}
which are consistent with the ER flows \eqref{Eq: ER-sq-Z3-A} and \eqref{Eq: ER-sq-Z3-B}, respectively.

\subsection{On the triangular and honeycomb lattice\label{subsection: ER_tri/hc}}
Next we move on to the triangular Laplacian model. Again, we first consider the $\Z_2$ case. The $\Z_2$ triangular Laplacian model is given by the stabilizer matrix 
\begin{equation}
    \A_\mathrm{tri}^{\Z_2}=\begin{pmatrix}
        x^2y+xy^2+x+y+x^2+y^2 
        \\ 
        \hline
        0
    \end{pmatrix}.
    \label{Eq: Z2tri}
\end{equation}
We first coarse-grain the model along $x$ and $y$ directions by $2$ and then apply local unitary transformations. As in \eqref{Eq: A_sq=2A'_sq}, it leads to two ``new'' models,
\begin{equation}
    \A_\mathrm{tri}^{\Z_2}(a)=\Tilde{\A}_\mathrm{1,tri}^{\Z_2}(2a) + \Tilde{\A}_\mathrm{2,tri}^{\Z_2}(2a),
    \label{Eq: A_tri=2A'_tri}
\end{equation}
where $=$ again means that we do not drop any disentangled qubits. The two new models are described by the stabilizer matrices
\begin{equation}
\begin{aligned}
    &\Tilde{\A}_\mathrm{1,tri}^{\Z_2}=\begin{pmatrix}
        x+y & 0 \\
        x+xy & 1+x+y+xy \\
        \hline
        \multicolumn{2}{c}{0}
    \end{pmatrix},
    \\
    &\Tilde{\A}_\mathrm{2,tri}^{\Z_2}=\begin{pmatrix}
        x+y & x+xy \\
        0 & 1+x+y+xy \\
        \hline
        \multicolumn{2}{c}{0}
    \end{pmatrix}.
    \label{Eq: tilde_A_Z2_tri}
\end{aligned}
\end{equation}
Furthermore, $\Tilde{\A}_{i,\mathrm{tri}}^{\Z_2}$ for $i=1,2$ is a quotient bifurcating fixed point. Explicitly,
\begin{equation}
    \Tilde{\A}_{i,\mathrm{tri}}^{\Z_2}(a) \xrightarrow{2\times 2} (\Tilde{\A}_{i,\mathrm{tri}}^{\Z_2} +\mathrm{Is}_{1+x}^{\Z_2}+\mathrm{Is}_{1+y}^{\Z_2}+\mathrm{Is}_{x+y}^{\Z_2})(2a),
    \label{Eq: ERG_A'_tri}
\end{equation}
where $i=1,2$ and $\mathrm{Is}^{\Z_2}$ represents a stack of 1D Ising models as in \eqref{Eq: ERG_A'_sq}.
The result resembles that of the $\Z_2$ square Laplacian model. Again, combining \eqref{Eq: A_tri=2A'_tri} and \eqref{Eq: ERG_A'_tri}, we identify that the $\Z_2$ triangular Laplacian model is a quotient bifurcating fixed point with the ER flow
\begin{equation}
    \A^{\Z_2}_\mathrm{tri}(a) \xrightarrow{4\times 4} \A^{\Z_2}_\mathrm{tri}(2a) + 2 (\mathrm{Is}_x^{\Z_2}+\mathrm{Is}_y^{\Z_2}+\mathrm{Is}_{x+y}^{\Z_2})(4a).
    \label{Eq: ERG_A_tri}
\end{equation}
After the anisotropic extension, we identify that the $\Z_2$ triangular Laplacian models bifurcates into itself and stacks of toric codes.

Again, we compare the GSDs of the models on both sides. Recall that the GSD of the $\Z_2$ triangular Laplacian model is given by \eqref{Eq: log2 GSD tri} (except for the factor of $2$, which came from anisotropic extension). It is easy to check that
\begin{equation}
\begin{aligned}
    k_\mathrm{tri}^{\Z_2}(4L_x,4L_y)
    &=k_\mathrm{tri}^{\Z_2}(2L_x,2L_y)+2 k_x^{\Z_2}(L_x,L_y) \\
    &\quad +2 k_y^{\Z_2}(L_x,L_y)+2 k_{x+y}^{\Z_2}(L_x,L_y)
\end{aligned}
\end{equation}
where the number of 1D Ising models in the stack described by $\mathrm{Is}_{x+y}^{\Z_2}$  is $\gcd(L_x,L_y)$. This is consistent with the ER flow \eqref{Eq: ERG_A_tri}.

When $N=3$, the stabilizer matrix takes the same form as \eqref{Eq: Z2tri} but the coefficients should be understood as modulo $3$. Similarly, the model seems non-bifurcating at first glance:
\begin{equation}
    \A_\mathrm{tri}^{\Z_3}(a) \xrightarrow{3\times 3}\mathrm{FSL}^{\Z_3}(3a) +\overline{\mathrm{FSL}}^{\Z_3}(3a) + 2 \tilde{\A}_\mathrm{tri}^{\Z_3}(3a),
    \label{Eq: Z_3_A_tri_ER}
\end{equation}
where $\mathrm{FSL}^{\Z_3}=(1+x+y,\;0)^\mathsf{T}$ is the $\Z_3$ fractal spin liquid, whose anisotropic extension was discussed in \eqref{Eq: fsl-stab-mat}, and
\begin{align}
    &\tilde{\A}_\mathrm{tri}^{\Z_3}=\begin{pmatrix}
        1+x+y & 0 \\
        1-y & x+y+xy \\
        \hline
        \multicolumn{2}{c}{0}
    \end{pmatrix},
    \label{Eq: B_tri}    
\end{align}
is a ``new'' model, analogous to \eqref{Eq: Z2-sq-newmodel} and \eqref{Eq: tilde_A_Z2_tri}.
One step forward, we show that the ERG flow of $\tilde{\A}_\mathrm{tri}^{\Z_3}$ is
\begin{equation}
    \tilde{\A}_\mathrm{tri}^{\Z_3}(a) \xrightarrow{3\times 3}\tilde{\A}_\mathrm{tri}^{\Z_3}(3a)+ 2 (\mathrm{FSL}^{\Z_3}+\overline{\mathrm{FSL}}^{\Z_3})(3a).\label{Eq: Z_3_B_tri_ER}
\end{equation}
And the $\Z_3$ fractal spin liquid is a self-bifurcating fixed point:
\begin{equation}
    \mathrm{FSL}^{\Z_3}(a) \xrightarrow{3\times 3} 3 \mathrm{FSL}^{\Z_3}(3a).
    \label{Eq: Z_3_FLS_ER}
\end{equation}
Following the same recipe as the $\Z_2$ Laplacian model on the square and triangular lattices, we find that the triangular $\Z_3$ Laplacian model is a quotient bifurcating fixed point by substituting \eqref{Eq: Z_3_B_tri_ER} and \eqref{Eq: Z_3_FLS_ER} into \eqref{Eq: Z_3_A_tri_ER}, i.e.,
\begin{equation}\label{Eq: ERG_A-tri-final}
    \A^{\Z_3}_\mathrm{tri}(a) \xrightarrow{9\times 9} \A^{\Z_3}_\mathrm{tri}(3a) + 6(\mathrm{FSL}^{\Z_3}+\overline{\mathrm{FSL}}^{\Z_3})(9a).
\end{equation}

Let us verify this ERG flow by comparing the GSDs on both sides. The GSD of the $\Z_3$ fractal spin liquid is given by $k_\mathrm{FSL}^{\Z_3}=3^{\min(k_x,k_y)}$, where $k_x$ and $k_y$ are defined as in \eqref{Eq: logpGSD_tri} (see Appendix \ref{appendix: gsd_fsl} for details).
Therefore, for these special values of $L_x$ and $L_y$, we have
\begin{equation}
\begin{aligned}
    k^{\Z_3}_\mathrm{tri}(9L_x,9L_y)
    &= k^{\Z_3}_\mathrm{tri}(3L_x,3L_y) + 6 k^{\Z_3}_\mathrm{FSL}(L_x,L_y) \\
    & \quad + 6 k^{\Z_3}_{\overline{\mathrm{FSL}}}(L_x,L_y),
\end{aligned}
\end{equation}
which agrees with the ERG flow \eqref{Eq: ERG_A-tri-final}.

Since the honeycomb Laplacian model equals to the triangular Laplacian model when $N=2$ and two copies of the fractal spin liquid when $N=3$, we can straightforwardly infer the ER of the honeycomb Laplacian model from the results above.

\section{Discussion and outlook\label{Sec: discussion}}

In this paper, we analyzed representative examples of anisotropic $\Z_N$ Laplacian models, defined on anisotropic extensions of various 2D regular lattices. 
Similar to the square lattice case \cite{PhysRevB.107.125121}, we found the scaling behaviour of the GSD and the mobility of point-like excitations to depend sensitively on the value of $N$. In addition, many features of fracton phases are, not surprisingly, sensitive to lattice geometries.
For example, the Kagome-based anisotropic $\Z_{2,3}$ Laplacian model is not robust while those based on the square, triangular, and honeycomb lattices are. 
Moreover, the entanglement renormalization group flows also varies with geometries: the anisotropic $\Z_{2,3}$ square-based and triangular-based Laplacian models are quotient bifurcating fixed points, whereas the anisotropic $\Z_3$ honeycomb Laplacian model is self-bifurcating. Incidentally, we also showed that the triangular-based and honeycomb-based models are equivalent when $N$ is not a multiple of $3$. 

We remark that the ER property can be sensitive to coarse-graining factors. Let us first consider the stabilizer models hosting string operators. This includes topological quantum liquid phases such the 2D or 3D toric code, foliated fracton order, and any anisotropic extension of a classical model, such as the anisotropic Laplacian model. (As explained before, the anisotropic extension necessarily hosts string operators along the anisotropic direction, but it is model-dependent in the other directions.) The topological quantum liquid phases are scale-invariant, thus are fixed points under the ER with arbitrary integer $c$ \cite{PhysRevB.79.195123,PhysRevLett.100.070404}. Similarly, fracton models that can be constructed by coupled layers are bifurcating for any $c$ when coarse-grained along the direction perpendicular to those layers. For example, we can coarse-grain the X-cube model by a factor of 3 along $x$ direction and the resulting Hamiltonian is equivalent to the X-cube model plus two stacks of 2D toric codes on the $yz$ planes. However, for models discussed in this paper, which do not host string operators within the $xy$ plane, we should be careful about the choice of coarse-graining factors, similar to Haah's cubic codes $\mathrm{CC}_{1-4,7,8,10}$.

There are several immediate questions that arise from this work. It would be interesting to systematically work out the entanglement renormalization flow of Laplacian models for arbitrary $N$. While a rigorous proof is lacking, an example for $N=5$ shows similar bifurcating behaviors as that of the $N=2, 3$ case, indicating that the ER captures universal features of fracton phases. When $N$ is composite, we expect that under ER, more elementary $\mathbb{Z}_{\tilde{N}}$ models will further be generated where $N=c\tilde{N}$ with $c>1$ is an integer.
The current approaches in this work rely on techniques from commutative algebra and are most useful when $N$ is prime. We leave a more systematic examination of the general cases for future work.

A more general question in the context of ER is determining when a model is a bifurcating fixed point. In the case of $\Z_{2,3}$ Laplacian models, we have shown that the model is reproduced under ER when the coarse-graining factor is chosen to be $c=2$ or $3$, respectively, in both $x$ and $y$ directions. This suggests that the $\Z_p$ Laplacian model is bifurcating under ER when coarse-grained by $c=p$. In fact, we believe that the same is true for any stabilizer code based on qudits of prime dimension $p$. It would be nice to confirm this intuition, perhaps using techniques from commutative algebra. We leave the resolution of these questions to future work.

We also believe that the study of ER can be generalized to a broader range of fracton order such as the fermionic fracton order \cite{PhysRevResearch.2.023353,shirley2020fractonicorderemergentfermionic,PhysRevLett.132.016604}, but we leave a careful study of these models to the future. Lastly, we remark that complementary tensor network methods such as MERAs \cite{PhysRevLett.112.240502} and TEFRs \cite{PhysRevB.80.155131} which are more focused on states can also be helpful in understanding the renormalization properties of fracton phases.

\begin{acknowledgments}
We thank Daniel Bulmash, Meng Cheng, Arpit Dua, Ho Tat Lam, Shu-Heng Shao, and Nathanan Tantivasadakarn for helpful discussions. Z.-X. L. is grateful to Andreas Karch and Hao-Yu Sun for their hospitality at UT Austin. 
Y. X. is supported by the National Science Foundation under NSF Award Number DMR – 2308817.   P. G. is supported by the Simons Collaboration on Global Categorical Symmetries. Z.-X. L. is partially supported by the Simons Collaborations on Ultra-Quantum Matter, grant 651440  from the Simons Foundation.
\end{acknowledgments}

\appendix

\section{Review of anisotropic $\Z_N$ Laplacian model on the square lattice\label{Appendix: Laplacian_square}}
In this appendix, we review the 3+1D $\Z_N$ anisotropic Laplacian model on an $L_x\times L_y\times L_z$ cubic lattice with periodic boundary conditions in all three directions \cite{PhysRevB.107.125121}. The graph $\Gamma$ is taken to be a torus $C_{L_x}\times C_{L_y} $. The stabilizer matrix of the corresponding (classical) Laplacian model
is $(f,0)^{\mathsf T}$, where
\begin{equation}
    f(x,y) = x(y-1)^2+y(x-1)^2 \mod N.
    \label{Eq: anisotropic square Laplacian}
\end{equation}

\begin{figure}
    \centering
    \includegraphics[scale=0.9]{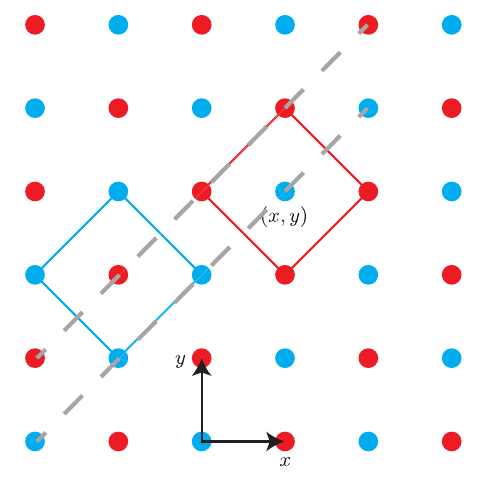}
    \caption{The anisotropic $\Z_2$ Laplacian model is equivalent to two copies of the anisotropic $\Z_2$ lineon model on the red and blue sublattices, respectively (the $z$-direction is suppressed). The gray dashed lines represent the string-like logical operators in the $(1,1)$ direction.}
    \label{fig: tilted}
\end{figure}

When $N=2$, assuming $L_x,L_y$ are both even or infinite, the square-based anisotropic Laplacian model is equivalent to two copies of the 3+1D anisotropic $\Z_2$ lineon model \cite{SHIRLEY2019167922,PhysRevB.103.205116} defined on a 
$45^\circ$ tilted square lattice \cite{PhysRevB.107.125121} shown in Fig. \ref{fig: tilted}. The logarithm of the GSD for $L_x=L_y=L$ is given by
\begin{equation}
    \log_2\mathrm{GSD}_\mathrm{sq}=
    \begin{cases}
        4L,& L\text{ is even,} \\
        4L-2,& L\text{ is odd.} \\
    \end{cases}
    \label{Eq: logGSD_Z2_square}
\end{equation}

When $N = p>2$ is an odd prime, things become significantly different. In this case, let $q\neq p$ be another odd prime such that $p$ is a primitive root modulo $q^m$ with integer $m\geq 1$. For $L_x=p^{k_x}q^m$ and $L_y=p^{k_y}q^m$ where $k_x,k_y,m,\geq 0$ are integers, we have
\begin{equation}
    \log_p\mathrm{GSD}_\mathrm{sq}=2(2p^{\min(k_x,k_y)}-\delta_{k_x,k_y}).
    \label{Eq: logp GSD sq}
\end{equation}

Let us discuss the mobility of the excitations on the infinite lattice. The string-like logical operators in the $z$-direction indicate that the model has lineons which can move along $z$-direction. However, the mobility of the $z$-lineons on the $xy$ plane is nontrivial. We will closely follow the argument in \cite{PhysRevB.107.125121} and Appendix \ref{appendix: mobility_tri_laplacian}.

For $N=2$, $f(x,y)$ in \eqref{Eq: anisotropic square Laplacian} is factorizable since $f(x,y)=(x+y)(1+\bar x \bar y)\mod 2$. To satisfy the mobility condition \eqref{Eq: mobility_condition}, we can take $(n_0,m_0)$ to be $(n,\pm n)$. When $(n_0,m_0)=(n,n)$, we have the factorization
\begin{equation}
    x^ny^n+1=(xy+1)(x^{n-1}y^{n-1}+\cdots+1)\mod 2.
\end{equation}
And $q(x,y)$ can be chosen as $x+y$. This is exactly the case that a dipole of $z$-lineons separated in the $(1,-1)$ direction can move in the $(1,1)$. The dashed lines in Fig. \ref{fig: tilted} represent the string operators that move the dipole. On the other hand, for $(n_0,m_0)=(n,-n)$, a dipole of $z$-lineons separated the in the $(1,1)$ direction can move in the $(1,-1)$. 

For $N=p>2$, $f(x,y)$ is irreducible up to any monomial. Following the argument in \cite{PhysRevB.107.125121}, we have that $x^{n_0}y^{m_0}-1$ is not a multiple of $f(x,y)$. Thus the condition \eqref{Eq: mobility_condition} cannot be satisfied, except trivially, and any finite set of $z$-lineons in the anisotropic $\Z_p$ Laplacian model cannot move in the $xy$ plane, except when they are created locally.

\onecolumngrid

\interfootnotelinepenalty=10000

\section{Ground-state degeneracies of the Laplacian models and the associated $\B$ codes \label{appendix: gsd_ca}}
\subsection{General procedure \label{subappendix: procedure}}
The analytical method to compute the ground-state degeneracy of translationally invariant Pauli stabilizer codes based on qudits of prime dimension $p$ involves using the Gr\"obner basis to calculate the dimension of a quotient ring, or more generally a quotient module \cite{Haah2013}\footnote{While \cite{Haah2013} considers only qubits explicitly, as stated there, all their results generalize straightforwardly to qudits of any prime dimension $p$.}. For concreteness, let us work in 2D with coordinates $(x,y)$ and periodic boundary conditions $(x,y) \sim (x+L_x,y) \sim (x,y+L_y)$. The starting point is the stabilizer map $\A : \F_p[x,y]^t \rightarrow \F_p[x,y]^{2q}$, which is given by the $2q\times t$ matrix defined in Sec. \ref{Section: Polynomial framwork}. In all our applications below, it turns out that $q=t$, i.e., the number of qudits per site is equal to the number of different type of stabilizer terms. In this case, the GSD is given by
\begin{equation}\label{Eq: GSD-general-expression}
    \log_p \mathrm{GSD} = \dim_{\F_p} \coker \A_L^\dagger = \dim_{\F_p} (\F_p[x,y]/\mathfrak b_L)^t/ \im \A_L^\dagger,
\end{equation}
where, $\mathfrak b_L = (x^{L_x}-1,y^{L_y}-1)$ is the ideal that effectively imposes the periodic boundary conditions and $\A_L$ is the same matrix as $\A$ that acts on the quotient ring $\F_p[x,y]/\mathfrak b_L$ instead of $\F_p[x,y]$.

For a classical code, only the first $q$ rows of $\A$ are nonzero, so we can replace $\A^\dagger$ with $(\A^Z)^\dagger$ in \eqref{Eq: GSD-general-expression}. Recall that the GSDs of a classical code and its anisotropic extension are related by \eqref{Eq: GSD-aniso-class}. So, in all applications below, we compute the $\log_p \mathrm{GSD}$ of the $\Z_p$ Laplacian models and their $\B$ codes, which are all classical codes. Multiplying the results by $2$ gives the expressions for the GSDs of their anisotropic extensions.

\subsubsection{Gr\"obner basis of ideals of polynomial rings for $q=t=1$}

In the special case of $q=t=1$, $\im \A^\dagger$ is simply the ideal $(f)$ of $\F_p[x,y]$, where $f(x,y)$ is the nonzero entry of $\A$\footnote{More precisely, $\im \A^\dagger$ is the ideal $(\bar f)$, but in our application, $\bar f(x,y) = f(\bar x,\bar y) = \bar x^2 \bar y^2 f(x,y)$ which is equivalent to $f(x,y)$ up to a monomial (i.e., translations).}. Therefore, the expression for GSD can be written as
\begin{equation}
    \log_p\mathrm{GSD} = \dim_{\F_p} \F_p[x,y]/\mathfrak i,
\end{equation}
where $\mathfrak i = (f) + \mathfrak b_L = (f,x^{L_x}-1,y^{L_y}-1)$.

Let us briefly review the Gr\"obner basis technique to compute this quantity. Given an ideal $\mathfrak i$ of the polynomial ring $\F[x,y]$, where $\F$ is a field, the vector space dimension of the quotient ring $\F[x,y]/\mathfrak i$ can be computed using the Gr\"obner basis of $\mathfrak i$. In particular, if $\G = \{g_1,\ldots,g_n\}$ denotes the Gr\"obner basis of $\mathfrak i$ with respect to a monomial order\footnote{We always use the lexicographic monomial order $x\succ y$ in computing the Gr\"obner basis.}, then any polynomial $P\in \F[x,y]$ can be written as
\begin{equation}
    P(x,y)=R(x,y)+\sum_{i=1}^n Q_i(x,y) g_i(x,y),
\end{equation}
where $R$ is not reducible with respect to $\G$\footnote{A polynomial $P$ is reducible with respect to a set of polynomials $\{g_1,\ldots,g_n\}$ if the leading term of $P$ is a multiple of the leading term of one of the $g_i$'s. Here, the leading term of a polynomial is defined with respect to the monomial order.}.
While $Q_i$'s (\`a la quotients) may not be uniquely determined, the polynomial $R$ (\`a la remainder) is uniquely determined by $P$. The uniquenss of $R$ is one of the defining properties of a Gr\"obner basis. Therefore, we may rewrite the above equation as
\begin{equation}
    P(x,y)=R(x,y)\mod \G.
\end{equation}
It follows that $\F[x,y]/\mathfrak i$ is in one-one correspondence with the set of polynomials that are irreducible with respect to $\G$. Therefore, the set of monomials that are irreducible with respect to $\G$ forms a basis of the vector space $\F[x,y]/\mathfrak i$. So, computing $\dim_\F \F[x,y]/\mathfrak i$ is equivalent to counting the number of monomials that are irreducible with respect to the Gr\"obner basis of $\mathfrak i$.

\subsubsection{Alternative expression for GSD when $q=t=1$}
While there are computer algebra systems that compute the Gr\"obner basis of an ideal quickly and efficiently, computing it analytically is hard. The following manipulations yield another expression for the GSD that is more amenable to analytic computation.

{It is well-known that any Artinian ring\footnote{An Artinian ring is a ring that satisfies the descending chain condition, i.e., if $\mathfrak{i}_1\supseteq \mathfrak{i}_2\supseteq\cdots$ is a descending chain of ideals, then there is a $k\ge 1$ such that $\mathfrak{i}_k = \mathfrak{i}_{k+1} = \cdots$. Any ring that is a finite dimensional vector space over a field is always Artinian, which is exactly the case here.} $\R$ 
has a finite number of maximal ideals and has a canonical decomposition $\R \cong \bigoplus_{\mathfrak m} \R_{\mathfrak m}$, where the sum is over the maximal ideals $\mathfrak m$ of $\R$ and $\R_{\mathfrak m}$ denotes the localization of $\R$ at $\mathfrak m$\footnote{The idea behind localization is to ``put in by hand'' multiplicative inverses of some elements in a ring $\R$. More precisely, the localization of $\R$ with respect to a multiplicatively closed set $S$ is a new ring, denoted as $S^{-1}\R$, whose elements are fractions with numerators in $\R$ and denominators in $S$. Given a maximal ideal $\mathfrak m$ of a ring $\R$, $\R\smallsetminus\mathfrak m$ is always multiplicatively closed. We define the localization of $\R$ at $\mathfrak m$, denoted as $\R_{\mathfrak m}$, as the localization of $\R$ with respect to $\R\smallsetminus \mathfrak m$.}. Therefore,
\begin{equation}
    \log_p \mathrm{GSD} = \dim_{\F_p} \F_p[x,y]/\mathfrak i= \sum_{\mathfrak m}\dim_{\F_p} (\F_p[x,y]/\mathfrak i)_{\mathfrak m}.
\end{equation}
The maximal ideals of $\F_p[x,y]/\mathfrak i$ are related to those of $\F_p[x,y]$ by the following theorem:
\begin{theorem}
    The maximal ideals of $\R/\mathfrak i$ are in one-one correspondence with the maximal ideals of $\R$ that contain $\mathfrak i$.
\end{theorem}

However, the maximal ideals of $\F_p[x,y]$ are hard to compute in general. Instead, it is simpler to work with the algebraic closure of $\F_p$, denoted as $\F = \overline{\F_p}$. Since the vector space dimension does not change under the extension of the underlying field, we have
\begin{equation}
    \log_p \mathrm{GSD}=\sum_{\mathfrak m}\dim_{\F} (\F[x,y]/\mathfrak i)_{\mathfrak m},
\end{equation}
where $\mathfrak i$ is now interpreted as the ideal of $\F[x,y]$ generated by the same polynomials as before.}

The simplicity of working with algebraic closure stems from the following theorem:
\begin{theorem}[Weak form of Hilbert's Nullstellensatz]
    Let $\F$ be an algebraically closed field. The maximal ideals of the polynomial ring $\F[x_1,\ldots,x_n]$ are of the form $(x_1-a_1,\ldots,x_n-a_n)$ for some $a_i\in \F$.
\end{theorem}
The maximal ideals of $\F[x,y]$ are therefore of the form $(x-x_0,y-y_0)$ for some $x_0,y_0\in \F$. The ideal $(x-x_0,y-y_0)$ contains $\mathfrak i$ if and only if $(x_0,y_0)$ is a root of all polynomials in $\mathfrak i$. Hence,
\begin{equation}
    \log_p \mathrm{GSD}=\sum_{(x_0,y_0)\in V(\mathfrak i)}\dim_{\F} (\F[x,y]/\mathfrak i)_{(x-x_0,y-y_0)},
\end{equation}
where 
\begin{equation}
    V(\mathfrak i)=\{(x,y)\in \F^2: f(x,y)=x^{L_x}-1=y^{L_y}-1=0 \}.
\end{equation}

To further simplify the calculations, let $L_i=p^{k_i}L_i'$ with $p\nmid L_i'$\footnote{$a\nmid b$ means $a$ does not divide $b$.}. Then we have the factorization,
\begin{equation}
\begin{aligned}
    x^{L_x}-1
    & =x^{L_x}-x_0^{L_x}=(x^{L'})^{p^{k_x}}-(x_0^{L'})^{p^{k_x}}=(x^{L'}-x^{L'}_0)^{p^{k_x}} \\
    &=(x-x_0)^{p^{k_x}}(x^{L'-1}+x^{L'-2}x_0+\cdots+x^{L'-1}_0)^{p^{k_x}},
\end{aligned}
\end{equation}
and similarly for $y^{L_y}-1$. The polynomial in the second bracket of the second line is a unit\footnote{A unit is an element of the ring with a mutliplicative inverse, i.e., $u$ is a unit if there exists $v$ in the ring such that $uv=1$.} in the localization $\F[x,y]_{(x-x_0,y-y_0)}$. Therefore, $(x^{L_x}-1,y^{L_y}-1)$ generates the same ideal as $((x-x_0)^{p^{k_x}},(y-y_0)^{p^{k_y}})$ in $\F[x,y]_{(x-x_0,y-y_0)}$. Defining 
\begin{equation}
    \mathfrak i_{x_0,y_0}=(f,(x-x_0)^{p^{k_x}},(y-y_0)^{p^{k_y}}),
\end{equation}
the quotient ring becomes\footnote{Given an ideal $\mathfrak i$ of a ring $\R$, the ideal $\mathfrak i_{\mathfrak m}$ of $\R_{\mathfrak m}$ is generated by the image of $\mathfrak i$ under the localization map $\R \rightarrow \R_{\mathfrak m}$.}
\begin{equation}
\begin{aligned}
    (\F[x,y]/\mathfrak i)_{(x-x_0,y-y_0)}
     &\cong  \F[x,y]_{(x-x_0,y-y_0)}/\mathfrak i_{(x-x_0,y-y_0)} \\
     & \cong  \F[x,y]_{(x-x_0,y-y_0)}/(\mathfrak i_{x_0,y_0})_{(x-x_0,y-y_0)} \\
     & \cong  (\F[x,y]/\mathfrak i_{x_0,y_0})_{(x-x_0,y-y_0)}.
\end{aligned}
\end{equation}
In the first and third lines, we used the fact that localization commutes with quotient. Note that the quotient ring $\F[x,y]/\mathfrak i_{x_0,y_0}$ is also Artinian, and by Hilbert's Nullstellensatz, its only maximal ideal is $(x-x_0,y-y_0)$. Therefore, 
\begin{equation}
    (\F[x,y]/\mathfrak i_{x_0,y_0})_{(x-x_0,y-y_0)}\cong \F[x,y]/\mathfrak i_{x_0,y_0},
\end{equation}
and hence,
\begin{equation}
    \log_p\text{GSD}=\sum_{(x_0,y_0)\in V(\mathfrak i)}\dim_{\F} \F[x,y]/\mathfrak i_{x_0,y_0}.
    \label{Eq: log_pGSD general}
\end{equation}
This is the desired alternative expression for the GSD.

\subsubsection{Gr\"obner basis of submodules of free modules for $q=t>1$}
When $q=t>1$, $\im \A^\dagger$ is not an ideal of $\F_p[x,y]$. Instead, it is a submodule of the free module $\F_p[x,y]^t$. The above Gr\"obner basis techniques generalize straightforwardly to this case. Let us briefly introduce the Gr\"obner basis of a submodule of a free module \cite{Lezama+2008+121+137}.

Let $\R=\F[x,y]$ be a polynomial ring over a field $\F$. Then, the Cartesian product 
\begin{equation}
    \R^t=\left\{ \vec f =(f_1,\ldots, f_t)^{\mathsf T} : f_i\in \R,\; i=1,\ldots,t \right\},
\end{equation}
is a free $\R$-module of rank $t$, i.e., it has a linearly independent finite generating set (i.e., a basis) of size $t$ given by
\begin{equation}
    \vec{e}_1=(1,0,\ldots,0)^\mathsf{T},\quad\ldots\quad,\quad\vec{e}_t=(0, \ldots,0,1)^\mathsf{T}.
\end{equation}
A submodule $\M$ of $\R^t$ is a subset that satisfies $r\cdot \vec f \in \M$ for any $r\in \R$ and $\vec f \in \M$. For example, given a $t\times s$ matrix $\tau: \R^s \rightarrow \R^t$, its image $\im \tau$ is a submodule of $\R^t$ generated by the columns of $\tau$. Note that this submodule is not necessarily a free $\R$-module, even though it has finitely many ($s$) generators because they can have nontrivial relations (known as ``syzygies'') among them.

A monomial of $\R^t$ is defined as a monomial of $\R$ times a basis vector, i.e., $x^a y^b \vec e_i$ for some nonnegative integers $a,b$ and $i=1,\ldots,t$. We say a monomial $x^a y^b \vec e_i$ is a multiple of another monomial $x^c y^d \vec e_j$ if $i=j$, $a\ge c$, and $b\ge d$. We define $\lcm(x^a y^b \vec e_i, x^c y^d \vec e_j) = x^{\max(a,c)} y^{\max(b,d)} \vec e_i$ if $i=j$ and $0$ otherwise.

Given a monomial order on $\R$, such as the lexicographic monomial order $x\succ y$, one can define several natural monomial orders on $\R^t$.
The one we will use is the ``term over position'' (TOP) order, where
\begin{equation}
    x^a y^b \vec e_i \succ x^c y^d \vec e_j \iff x^a y^b \succ x^c y^d, \quad \text{or} \quad x^a y^b = x^c y^d \text{ and } i>j.
\end{equation}
The leading term of $\vec f \in \R^t$, denoted as $\mathrm{lt}(\vec f)$, is the largest term in $\vec f$ with respect to the monomial order. For any nonempty subset $\mathcal S$ of $\R^t$, we define $\mathrm{lt}(\mathcal S)$ as the submodule of $\R^t$ generated by the leading terms of the elements of $\mathcal S$.

We now give the definition of the Gr\"obner basis for a submodule of a free $\R$-module \cite{Lezama+2008+121+137}, which is quite similar to that of an ideal of $\R$. 
\begin{definition}
     Let $\M\neq 0$ be a submodule of $\R^t$ and $\G$ be a finite subset of $\M$. Then $\G$ is a Gr\"obner basis of $\M$ if $\mathrm{lt}(\M)=\mathrm{lt}(\G)$.
\end{definition}
\begin{theorem}
    Let $\M\neq 0$ be a submodule of $\R^t$ and $\G=\{\vec{g}_1,\ldots,\vec{g}_n\}$ be a finite subset of nonzero vectors of $\M$. Then the following are equivalent:
    \begin{enumerate}[(i)]
        \item $\G$ is a Gr\"obner basis of $\M$.
        \item For any vector $\vec{f}\in \R^t$, $\vec{f}\in \M\iff $ $\vec{f}$ can be reduced to $\vec{0}$ by $\G$.\footnote{A vector $\vec f\in \R^t$ is reducible with respect to a set of vectors $\{\vec g_1,\ldots,\vec g_n\}$ if the leading term of $\vec f$ is a multiple of the leading term of one of the $\vec g_i$'s.}
        \item For any vector $\vec{f}\in \M$, there exists polynomials $q_1,\ldots,q_n\in \R$ such that
        \begin{equation*}
            \vec{f}=\sum_{i=1}^n q_i\vec{g}_i.
        \end{equation*}
    \end{enumerate}
\end{theorem}

Similar to the case of a quotient ring, the quotient module 
$\R^t/\M$ is a set of equivalence classes, each of which is represented by a vector in $\R^t$ that is irreducible with respect to the Gr\"obner basis $\G$ of $\M$. Therefore, the vector-space dimension of the quotient module, $\dim_{\F} (\R^t/\M)$, is equal to the number of monomials of $\R^t$ that are irreducible with respect to $\G$.

\subsubsection{Alternative expression for GSD when $q=t>1$}
As in the case of ideals, computing the Gr\"obner basis of submodules of free modules over $\F_p[x,y]$ analytically is very hard. So, once again, we perform the following manipulations to get another expression for the GSD that is more tractable analytically.

The first step is to localize the module at the maximal ideals of $\F_p[x,y]$. Say $\R$ is an Artinian ring, which means it has a finite number of maximal ideals and satisfies $\R \cong \bigoplus_{\mathfrak m} \R_{\mathfrak m}$. Then, any finitely generated $\R$-module $\M$ satisfies $\M \cong \bigoplus_{\mathfrak m} \M_{\mathfrak m}$\footnote{To see this, first tensor both sides of $\R \cong \bigoplus_{\mathfrak m} \R_{\mathfrak m}$ with $\M$ and then use the facts that $\R \otimes_\R \M \cong \M$, tensoring is right-exact, and tensoring commutes with direct sum and localization.}, where $\M_{\mathfrak m}$ denotes the localization of $\M$ at $\mathfrak m$. Moreover, $\M_{\mathfrak m}$ is nonzero if and only if $\mathfrak m$ contains the zeroth Fitting ideal of $\M$. If $\tau: \R^s \rightarrow \R^t$ is a $t\times s$ matrix, then the $k$-th Fitting ideal of $\M = \coker \tau$ is the ideal generated by the $(t-k)\times (t-k)$ minors of $\tau$. Using these facts for $\R = \F_p[x,y]/\mathfrak b_L$ and $\tau = \A^\dagger$, the GSD is given by
\begin{equation}
    \log_p\mathrm{GSD} =  \sum_{\mathfrak m \supseteq (f)} \dim_{\F_p} [(\F_p[x,y]/\mathfrak b_L)^t/\im \A^\dagger]_{\mathfrak m},
\end{equation}
where $f(x,y)=\det (\A^Z)^\dagger$ is the only nonzero $t\times t$ minor of $\A^\dagger$.

Next, we extend the base field from $\F_p$ to its algebraic closure $\F = \overline{\F_p}$. This does not change the vector-space dimension, so
\begin{equation}
    \log_p\mathrm{GSD} = \sum_{\mathfrak m\supseteq (f)} \dim_\F [(\F[x,y]/\mathfrak b_L)^t/\im \A^\dagger]_{\mathfrak m}.
\end{equation}
The maximal ideals of $\F[x,y]/\mathfrak b_L$ are in one-one correspondence with maximal ideals of $\F[x,y]$ that contain $\mathfrak b_L$. By Hilbert's Nullstellensatz, the maximal ideals of $\F[x,y]$ are of the form $(x-x_0,y-y_0)$ for some $x_0,y_0\in \F$. And the ideal $(x-x_0,y-y_0)$ contains $\mathfrak i$ if and only if $(x_0,y_0)$ is a root of all the polynomials in $\mathfrak i$. Hence,
\begin{equation}
    \log_p \mathrm{GSD} = \sum_{(x_0,y_0) \in V(\mathfrak i)} \dim_\F [(\F[x,y]/\mathfrak b_L)^t/\im \A^\dagger]_{(x-x_0,y-y_0)},
\end{equation}
where $\mathfrak i = (f) + \mathfrak b_L$ and
\begin{equation}
    V(\mathfrak i) = \{(x,y)\in \F^2:f(x,y) = x^{L_x}-1 = y^{L_y}-1 = 0\}.
\end{equation}

Let $L_i = p^{k_i} L_i'$, where $p\nmid L_i'$. By the same argument as before, within the localization at $(x-x_0,y-y_0)$, we can replace $x^{L_x}-1$ with $(x-x_0)^{p^{k_x}}$ and $y^{L_y}-1$ with $(y-y_0)^{p^{k_y}}$. That is, $\mathfrak b_{L;x_0,y_0} = ((x-x_0)^{p^{k_x}},(y-y_0)^{p^{k_y}})$ is the same ideal as $\mathfrak b_L$ in $\F[x,y]_{(x-x_0,y-y_0)}$. Therefore, we have
\begin{equation}
    \log_p \mathrm{GSD} = \sum_{(x_0,y_0) \in V(\mathfrak i)} \dim_\F [(\F[x,y]/\mathfrak b_{L;x_0,y_0})^t/\im \A^\dagger]_{(x-x_0,y-y_0)}.
\end{equation}

Now, $\F[x,y]/\mathfrak b_{L;x_0,y_0}$ is an Artinian ring, and by Hilbert's Nullstellensatz, its only maximal ideal is $(x-x_0,y-y_0)$, so $(\F[x,y]/\mathfrak b_{L;x_0,y_0})_{(x-x_0,y-y_0)} \cong \F[x,y]/\mathfrak b_{L;x_0,y_0}$. Therefore,
\begin{equation}
    \begin{aligned}\label{Eq: log_p GSD t>1}
        \log_p \mathrm{GSD} = \sum_{(x_0,y_0) \in V(\mathfrak i)} \dim_\F (\F[x,y]/\mathfrak b_{L;x_0,y_0})^t/\im \A^\dagger = \sum_{(x_0,y_0) \in V(\mathfrak i)} \dim_\F \F[x,y]^t/\im \tau_{x_0,y_0},
    \end{aligned}
\end{equation}
where
\begin{equation}
    \tau_{x_0,y_0} = \begin{pmatrix}
        \A^\dagger & (x-x_0)^{p^{k_x}} \vec 1_t & (y-y_0)^{p^{k_y}} \vec 1_t
    \end{pmatrix}.
\end{equation}
It is easy to check that when $q=t=1$, this expression reduces to the one in \eqref{Eq: log_pGSD general}.

\subsection{$\Z_p$ Laplacian model on the triangular lattice\label{appendix: Zp_tri_gsd}}
In this case, $q=t=1$, so we use the alternative expression \eqref{Eq: log_pGSD general} to compute the GSD of this model for some special values of $L_x$ and $L_y$. The ideal we are dealing with is $\mathfrak i=(f,x^{L_x}-1,y^{L_y}-1)$ with $f(x,y)=-6 xy+x+y+x^2+y^2+x^2y+xy^2$. Let $(x_0,y_0)$ be the solution of the system of polynomial equations
\begin{equation}
    f(x,y)=x^{L_x}-1=y^{L_y}-1=0.
\end{equation}
Under the factorization $x^{L_x}-1=(x^{L_x'}-1)^{p^{k_x}}$ where $p\nmid L_x'$, it becomes
\begin{equation}
    f(x,y)=x^{L_x'}-1=y^{L_y'}-1=0.
    \label{Eq: variety}
\end{equation}

Let us start from the simplest special case $L_x'=L_y'=1$. Obviously, the only solution in this case is $(x_0,y_0)=(1,1)$. 
We assume without loss of generality that $k_x\geq k_y$. The Gr\"obner basis for $p=3$ and $p>3$ are different, so we discuss them separately. 

For $p=3$, $f(x,y) = (1+x+y)(x+y+xy)$, and the Gr\"obner basis is given as
\begin{equation}
    g_1 = y^{3^{k_y}}-1,\qquad g_2 = x^2+x\sum_{i=2}^{3^{k_y}-1}(-y)^i+y.
\end{equation}
The set of monomials that are irreducible with respect to this Gr\"obner basis is
\begin{equation}
    \{1,y,\ldots,y^{3^{k_y}-1},x,xy,\ldots,xy^{3^{k_y}-1}\},
\end{equation}
with cardinality $2\times 3^{k_y}$. By \eqref{Eq: log_pGSD general}, we conclude
\begin{equation}
    \log_3\mathrm{GSD}_\mathrm{tri}=2\times 3^{k_y}.
    \label{Eq: log_3_GSD_tri}
\end{equation}

For $p>3$, the Gr\"obner basis is given as follows:
\begin{enumerate}[(1)]
    \item When $k_x\neq k_y$, 
    \begin{equation}
        g_1 = y^{p^{k_y}}-1,\qquad g_2 = x^2+4x\sum_{i=2}^{p^{k_y}-1} (-y)^i-3xy-3x+y.
    \end{equation}
    The set of monomials which are irreducible with respect to this Gr\"obner basis is
    \begin{equation}
        \{1,y,\ldots,y^{p^{k_y}-1},x,xy,\ldots,xy^{p^{k_y}-1}\},
    \end{equation}
    with cardinality $2 p^{k_y}$.
    \item When $k_x=k_y=k$\footnote{Note that this Gr\"obner basis is minimal but not reduced, i.e., we can reduce the subleading terms of $g_2$ with respect to $g_3$, but we choose not to do so. This does not affect any of our conclusions.},
    \begin{equation}
        g_1 = y^{p^{k_y}}-1,\qquad g_2 = x^2+4x\sum_{i=2}^{p^{k_y}-1} (-y)^i-3xy-3x+y,\qquad g_3 = (x-1)\sum_{i=0}^{p^k-1}y^i.
    \end{equation}
    The set of monomials which are irreducible with respect to this Gr\"obner basis is
    \begin{equation}
        \{1,y,\ldots,y^{p^{k_y}-1},x,xy,\ldots,xy^{p^{k_y}-2}\},
    \end{equation}
    with cardinality $2 p^{k_y}-1$.
    
\end{enumerate}
Combining the two cases together, the number of irreducible monomials is $2p^{k_y}-\delta_{k_x,k_y}$. By \eqref{Eq: log_pGSD general}, we conclude
\begin{equation}
    \log_p\mathrm{GSD}_\mathrm{tri}=2 p^{k_y}-\delta_{k_x,k_y},\quad p>3.
    \label{Eq: log_p_GSD_tri}
\end{equation}

\begin{proof}[Proof of Gr\"obner basis]
    Let us work with arbitrary odd prime $p$, including $p=3$, and deal with the special cases when necessary. It is easy to check that each of the above Gr\"obner bases satisfies the Buchberger's criterion. It is also easy to verify that
    \begin{equation}
        f = -4x g_1 + (y+1) g_2,\qquad g_2 = a\left[ \sum_{i=0}^{p^{k_y}-1} (-y)^i  f + (x^2 + xy -7x +y) (y^{p^{k_y}}-1) \right],
    \end{equation}
    where $a$ is the reciprocal of $2$ modulo $p$, i.e., $a$ is an integer satisfying $2a = 1 \mod p$, which exists because $p$ is an odd prime. This shows that $f \in (g_1,g_2)$ and that $g_2 \in \mathfrak i$. We only have to take care of $x^{p^{k_x}}-1$ and $g_3$ now.
    
    It is convenient to work with the variables $u=x-1$ and $v=y-1$, with lexicographic monomial order $u\succ v$. In terms of these variables, the ideal is $\mathfrak i = (\tilde f,u^{p^{k_x}},v^{p^{k_y}})$, where $\tilde f(u,v) = f(u+1,v+1)$, and we have
    \begin{equation}
        \tilde g_1 = v^{3^{k_y}},\qquad \tilde g_2 = u^2 + u \left(-v+4\sum_{i=2}^{p^{k_y}-1} (-av)^i \right) + 4\sum_{i=2}^{p^{k_y}-1} (-av)^i,\qquad \tilde g_3 = u v^{p^{k_y}-1}.
    \end{equation}
    where, once again, $a$ is the reciprocal of $2$ modulo $p$, and $\tilde g_3$ is needed only when $p>3$ and $k_x =  k_y$ as we will show below. 
    
    Let us show that $u^{p^{k_x}}$ is contained in the ideal generated by the Gr\"obner basis. Consider a polynomial of the form
    \begin{equation}
        s = \sum_{i=0}^k u^{k-i} v^i s_i(v),
    \end{equation}
    where $s_i(v)$ are polynomials in $v$ independent of $u$. For example, $u^{p^{k_x}}$ and $\tilde g_2 = u^2 + u v h_1(v) + v^2 h_2(v)$ are both of this form, where $h_{1,2}(v)$ are polynomials in $v$. We want to reduce $s$ with respect to $\tilde g_2$ as much as possible. First, we have
    \begin{equation}
        s^{(1)} = s - s_0(v) u^{k-2} \tilde g_2 = u^{k-1} v [s_1(v) - s_0(v) h_1(v)] + u^{k-2} v^2 [s_2(v) - s_0(v) h_2(v)] + \sum_{i=3}^k u^{k-i} v^i s_i(v).
    \end{equation}
    Clearly, $s^{(1)}$ is also of the same form as $s$ but the leading term of $s^{(1)}$ is smaller than that of $s$. Repeating this $k-1$ times, we end up with the polynomial $s^{(k-1)}(u,v)$ of the form $u v^{k-1} s^{(k-1)}_{k-1}(v) + v^k s^{(k-1)}_k(v)$, which is not reducible with respect to $\tilde g_2$ anymore.
    
    Applying this procedure to $u^{p^{k_x}}$, we get a polynomial of the form $u v^{p^{k_x}-1} r_1(v) + v^{p^{k_x}} r_2(v)$. When $k_x > k_y$, both terms are reducible with respect to $\tilde g_1$, which means $u^{p^{k_x}}$ is in the ideal generated by $\tilde g_1$ and $\tilde g_2$, and we are done. In particular, we do not need to add $\tilde g_3$ to the Gr\"obner basis.
    
    On the other hand, when $k_x = k_y$, things are a bit more subtle. The second term is still reducible with respect to $\tilde g_1$, but the first term is reducible with respect to $\tilde g_1$ if and only if $r_1(0) = 0$, i.e., the constant term of $r_1(v)$ vanishes (the non-constant terms are reducible with respect to $\tilde g_1$). If $r_1(0) \ne 0$, however, then we need $\tilde g_3$ to reduce the first term to $0$. We will show that $r_1(0) = 0$ only when $p=3$.

    Let $\hat g_2 = u^2 + uv h_1(0) + v^2 h_2(0) = u^2 - uv + v^2 = (u-av)^2 + b v^2 \mod p$, where $b=1-a^2 = 3a^2 \mod p$. Repeating the above reduction procedure on $u^{p^{k_y}}$ with respect to $\hat g_2$, we again get a polynomial of the form $u v^{p^{k_y}-1} \hat r_1(v) + v^{p^{k_y}} \hat r_2(v)$. Moreover, it is clear that $\hat r_{1,2}(v)$ is obtained from $r_{1,2}(v)$ by replacing $h_{1,2}(v)$ with $h_{1,2}(0)$. This means, $r_{1,2}(0) = \hat r_{1,2}(0)$. Now, consider the equation
    \begin{equation}
        u^{p^{k_y}} = (u-av)^{p^{k_y}} + (av)^{p^{k_y}} = (u-av)(\hat g_2 - bv^2)^{(p^{k_y}-1)/2} + (av)^{p^{k_y}}. 
    \end{equation}
    This means, after reducing $u^{p^{k_y}}$ with respect to $\hat g_2$, we get
    \begin{equation}
        (-b)^{(p^{k_y}-1)/2}(u-av)v^{p^{k_y}-1} - (av)^{p^{k_y}} \implies \hat r_1(v) = (-b)^{(p^{k_y}-1)/2}~.
    \end{equation}
    When $p=3$, we have $b=0 \implies \hat r_1(0) = 0 \implies r_1(0) = 0$, so $u^{3^{k_y}} \in (\tilde g_1,\tilde g_2)$, and there is no need to include $\tilde g_3$ in the Gr\"obner basis. On the other hand, when $p>3$, we have $b = 3a^2 \ne 0\mod p \implies \hat r_1(0) \ne 0 \implies r_1(0) \ne 0$, so
    \begin{equation}
        u^{p^{k_y}} = r_1(0) u v^{p^{k_y}-1} \mod (\tilde g_1,\tilde g_2).
    \end{equation}
    This additional term is reducible with respect to $\tilde g_3$, so $u^{p^{k_y}} \in (\tilde g_1,\tilde g_2,\tilde g_3)$.

    Finally, we need to show that $\tilde g_3 \in \mathfrak i$ when $k_x = k_y$ and $p>3$. Since $r_1(0)\ne 0 \mod p$ in this case, the above equation can be rewritten as
    \begin{equation}
        \tilde g_3 = r_1(0)^{-1}u^{p^{k_y}} \mod (\tilde f,v^{p^{k_y}}),
    \end{equation}
    where we used the fact that $\tilde g_1,\tilde g_2 \in (\tilde f,v^{p^{k_y}})$. Therefore, $\tilde g_3 \in \mathfrak i$, and we are done.
\end{proof}

Now we generalize the special case $L'_x=L'_y=1$ to $L_x'=L_y'=q$, where $q>2$ is a prime such that $p$ is a primitive root modulo $q$\footnote{$p$ is a primitive root modulo $q$ if for every integer $c$ coprime to $q$, there exists an integer $n$ such that $p^n=c\mod q$, or equivalently, $p$ is the generator of the multiplicative group of integers modulo $q$, denoted as $\Z^\times_q$.}. The equations \eqref{Eq: variety} become
\begin{equation}
    f(x,y)=x^{q}-1=y^q-1=0.
    \label{Eq: q}
\end{equation}
Let us prove that $x_0=y_0=1$ is the only solution by contradiction. Assume that there exists another solution $(x_0,y_0)$ with $x_0\neq 1$. Since $x_0^q=1$ and $x_0\neq 1$, powers of $x_0$ generate all the $q$-th roots of 1. Hence $y_0=x_0^s$ with $0<s<q$, where $s$ cannot take 0 since there is no solution of \eqref{Eq: q} of the form $(x_0,1)$ with $x_0\neq 1$. Furthermore, \eqref{Eq: q} is invariant under the transformation $y_0\leftrightarrow y_0^{-1}$. So $(x_0,x_0^{q-s})$ is also a solution. We can then obtain a solution of the form $(x_0,x_0^s)$ with $1\leq s\leq (q-1)/2$. Now $f(x,y)=0$ can be rewritten as $f(x,x^s)=0$ where $x_0$ is a solution of it. Explicitly,
\begin{equation}
    f(x,y) = x(y-1)^2 + y(x-1)^2 + (x-y)^2 \implies f(x,x^s)=x(x-1)^2\Tilde{f}_s(x),
    \label{Eq: f_s}
\end{equation}
where
\begin{equation}
    \Tilde{f}_s(x)=\left(\sum_{i=0}^{s-1} x^i\right)^2 + x^{s-1} + x\left(\sum_{i=0}^{s-2} x^i\right)^2.
\end{equation}
In $\Z_p[x]$, $\Tilde{f}_s(x)$ is nonzero because $\Tilde{f}_s(0)=1 \mod p$ for $s>1$ and $2 \mod p$ for $s=1$. Clearly, $x_0$ must be a solution of $\Tilde{f}_s(x)$ since $x_0\neq 0,1$. But $x_0$ is also a root of the cyclotomic polynomial $\Phi_q(x)=\sum_{i=0}^{q-1}x^i$. Now we use the fact that $\Phi_q(x$) must be a minimal polynomial of $x$ in $\Z_p[x]$ since $p$ is a primitive root modulo $q$ \cite[Section~11.2.B]{cox2012galois}. This means $\Tilde{f}_s(x)$ must be divisible by $\Phi_q(x)$. However,
\begin{equation}
    \deg_x\Tilde{f}_s(x)=2s-2\leq q-3<q-1=\deg_x \Phi_q(x),
\end{equation}
which is impossible, so there is no such $x_0$. Therefore, $x_0=y_0=1$ is the only solution in this case and $\log_p\mathrm{GSD_{tri}}$ is again given by \eqref{Eq: log_3_GSD_tri} and \eqref{Eq: log_p_GSD_tri}.

Finally, we further generalize the case to $L'_x=L'_y=q^m$, where $q>2$ is a prime and $m$ is a positive integer such that $p$ is a primitive root modulo $q^m$. Then, the equations \eqref{Eq: variety} become
\begin{equation}
    f(x,y)=x^{q^m}-1=y^{q^m}-1=0.
    \label{Eq: variety q^m}
\end{equation}
Again, we assume there exists the solution other than $(1,1)$ that can be written as $(x_0,x_0^s)$ for $1\leq s\leq (q^m-1)/2$ by a similar argument as before. The range of $s$ can be further narrowed down to $1\leq s\leq (q^r-1)/2$ where we assume there exists $0< r \leq m$ such that $x_0^{q^r}=1$ but $x_0^{q^{r'}}\neq 1$ for $r'<r$. We exclude the case $r=0$ since it corresponds to the solution $(1,1)$. According to the Euler totient function $\varphi(q^m)=q^m-q^{m-1}$, we separate the discussion into two cases:
\begin{enumerate}[(1)]
    \item $1\leq s\leq \varphi(q^r)/2$: Since $x_0^{q^r}=1$, $x_0$ is a root of the cyclotomic polynomial $\Phi_{q^r}(x)=\sum_{i=0}^{q-1}x^{i q^{r-1}}$. Meanwhile, since $x_0\neq 0,1$, it is also a root of $\Tilde{f}_s(x)$ in \eqref{Eq: f_s}. Again, we use the fact that $\Phi_{q^r}(x)$ is a minimal polynomial of $x_0$ in $\Z_p[x]$ \cite{doi:https://doi.org/10.1002/9781118218457.ch9}, so $\Phi_{q^r}(x)$ must divide $\Tilde{f}_s(x)$. But this is impossible since
    \begin{equation}
        \deg_x\Tilde{f}_s(x)=2s-2\leq \varphi(q^r)-2 < \varphi(q^r)=\deg_x\Phi_{q^r}(x).
    \end{equation}
    Therefore, such an $x_0$ does not exist.
    \item $\varphi(q^r)/2<s\leq (q^r-1)/2$: For convenience, let $t=q^r-2s$ so that the range of $t$ is $1\leq t < q^{r-1}\leq (q^r-q^{r-1})/2=\varphi(q^r)/2 $ where the rightmost inequality holds for $q>2$. There is a $z_0 \in \F$ such that $z_0^{q^r}=1$ but $z_0^{q^{r'}}\neq 1$ for $r'<r$, and $x_0 = z_0^2$ (which is possible because $q>2$). Then, the solution $(x_0,x_0^s)$ is of the form $(z_0^2,z_0^{2s})$. The transformation $y_0\leftrightarrow y_0^{-1}$ implies $(z_0^2,z_0^{t})$ is also a solution. It follows that $z_0$ is a root of
    \begin{equation}
        f(x^2,x^t) = \begin{cases}
            x(x-1)^2 \tilde f_1(x),&t=1,
            \\
            x^2(x-1)^2 \tilde f_t(x),&t>1,
        \end{cases} 
    \end{equation}
    where
    \begin{equation}
        \tilde f_t(x) = \begin{cases}
            x + (x+1)^2 + x,&t=1
            \\
            \left(\sum_{i=0}^{t-1} x^i\right)^2 + x^{t-2}(x+1)^2 + x^2\left(\sum_{i=0}^{t-3} x^i\right)^2,& t> 1,
        \end{cases}
    \end{equation}
    Note that $\Tilde{f}_t(x)$ is nonzero because $\Tilde{f}_t(0)=1\mod p$ for $t\neq 2$ and $2\mod p$ for $t=2$. Meanwhile, as $z_0$ is a root of $f(x^2,x^t)$ and $z_0\neq 0,1$, it is also a root of $\Tilde{f}_t(x)$. Again, we use the fact that $\Phi_{q^r}(x)$ is a minimal polynomial of $z_0$, so $\Phi_{q^r}(x)$ must divide $\Tilde{f}_t(x)$. But this is impossible because
    \begin{equation}
        \deg_x\Tilde{f}_t(x)=2t-2+2\delta_{t,1}\leq 2t < \varphi(q^r)=\deg_x\Phi_{q^r}(x).
    \end{equation}
    Thus, there is no such $z_0$.
    
\end{enumerate}
Therefore, when $p$ is a primitive root modulo $q^m$, $(1,1)$ is the only solution of \eqref{Eq: variety q^m}. Hence, $\log_p\mathrm{GSD}_\mathrm{tri}$ is still given by \eqref{Eq: log_3_GSD_tri} and \eqref{Eq: log_p_GSD_tri}.

To conclude, when $L_x=p^{k_x}q^m$ and $L_y=p^{k_y}q^m$, where $k_x,k_y,m\geq 0$ are integers and $q$ is an odd prime such that $p$ is a primitive root modulo $q^m$ (for $m\ge 1$), the ground-state of the triangular-based $\Z_p$ Laplacian model is given by
\begin{equation}
    \log_p\mathrm{GSD}_\mathrm{tri}=\left\{
    \begin{aligned}
        & 2\times 3^{\min(k_x,k_y)}, & p=3,\\
        & 2p^{\min(k_x,k_y)}-\delta_{k_x,k_y},  & p>3.
    \end{aligned}
    \right.
\end{equation}

\subsection{$\Z_3$ fractal spin liquid\label{appendix: gsd_fsl}}

Once again, $q=t=1$ in this case, so we use \eqref{Eq: log_pGSD general} to compute the GSD for some special values of $L_x$ and $L_y$. The ideal is $\mathfrak i = (f,x^{L_x}-1,y^{L_y}-1)$, where $f(x,y) = 1+x+y$. Like before, we consider the factorization $x^{L_x}-1=(x^{L_x'}-1)^{3^{k_x}} $,  $y^{L_y}-1=(y^{L_y'}-1)^{3^{k_y}} $
and take the special case $L_x'=L_y'=1$. The system of polynomial equations we need to solve is
\begin{equation}
    1+x+y=x-1=y-1=0.
    \label{Eq: variety fsl}
\end{equation}
Clearly, the only solution is $(1,1)$, so the ideal is $(1+x+y,x^{3^{k_x}}-1,y^{3^{k_y}}-1)$. Assuming without loss of generality that $k_x\geq k_y$,  the Gr\"obner basis is easily seen to be
\begin{equation}
    g_1=y^{3^{k_y}}-1,\qquad g_2 = x+y+1.
\end{equation}
The only slightly nontrivial fact to verify is that 
\begin{equation}
    x^{3^{k_x}}-1 = (g_2-y-1)^{3^{k_x}}-1 = g_2^{3^{k_x}}-(y^{3^{k_x}}-1) = 0 \mod (g_1,g_2).
\end{equation}
The set of monomials which are irreducible with respect to the Gr\"obner basis is
\begin{equation}
    \{1,y,\ldots,y^{3^{k_y}-1}\},
\end{equation}
with cardinality $3^{k_y}$. By \eqref{Eq: log_pGSD general}, \begin{equation}
    \log_3\mathrm{GSD}_\mathrm{FSL}= 3^{k_y}.
    \label{Eq: log_3_fsl}
\end{equation}

We use the same argument as before to generalize the simplest case $L_x'=L_y'=1$ to the more general case $L_x'=L_y'=q$ where $q$ is an odd prime such that $3$ is a primitive root modulo $q$ (e.g., $q=5$). Assume there exists a solution other than $(1,1)$ which can be written as $(x_0,x_0^s)$ with $x_0\neq1$ and $1\leq s\leq (q-1)/2$ by the same reasoning as in Appendix \ref{appendix: Zp_tri_gsd}. Then $x_0$ is a root of
\begin{equation}
    f(x,x^s)=1+x+x^s=(x-1)\Tilde{f}_s(x),\qquad \Tilde{f}_s(x)=1+\sum_{i=0}^{s-1}x^i.
    \label{Eq: f_fsl}
\end{equation}
Clearly, $\Tilde{f}_s(x)$ is nonzero since $\Tilde{f}_s(0)=-1\mod 3$. And since $x_0\neq1$, $x_0$ must be a solution of $\Tilde{f}_s(x)$. Again, the cyclotomic polynomial $\Phi_q(x)$ is a minimal polynomial of $x_0$ in $\Z_3[x]$, so it must divide $\Tilde{f}_s(x)$. But this is impossible since
\begin{equation}
    \deg\Tilde{f}_s(x)=s-1\leq \frac{q-3}{2}<q-1=\deg_x \Phi_q(x).
\end{equation}
So there is no such $x_0$, and hence $\log_3\mathrm{GSD}_\mathrm{FSL}$ is given by \eqref{Eq: log_3_fsl}.

Generalizing further, let $L_x'=L_y'=q^m$, where $m$ is a positive integer and $q$ is an odd prime such that $3$ is a primitive root modulo $q^m$ (e.g., $q=5$ and any $m\ge1$). Assume there exists a solution of the form $(x_0,x_0^s)\ne (1,1)$ for $1\leq s \leq (q^r-1)/2$ where $0<r\leq m$ and $x_0^{q^r}=1$ but $x_0^{q^{r'}}\neq 1$ for $r'<r$. 
\begin{enumerate}[(1)]
    \item $1 \leq s\leq \varphi(q^r)/2$: $x_0$ is a root of the cyclotomic polynomial $\Phi_{q^r}(x)$ and $\Tilde{f}_s(x)$ in \eqref{Eq: f_fsl}. But $\Phi_{q^r}(x)$ is a minimal polynomial of $x_0$ in $\Z_3[x]$ so it must divide $\Tilde{f}_s(x)$, which is impossible since
    \begin{equation}
        \deg\Tilde{f}_s(x)=\varphi(q^r)/2-1<\varphi(q^r)=\deg_x\Phi_{q^r}(x).
    \end{equation}
    So there is no such $x_0$. 
    \item $\varphi(q^r)/2<s<(q^r-1)/2$: Again, we let $t=q^r-2s$ so that we have $1\leq t<\varphi(q^r)/2$. By the same reasoning as before, we can write $x_0 = z_0^2$ and so $(z_0^2,z_0^t)$ is a solution. Thus $z_0$ is a root of
    \begin{equation}
        f(x^2,x^t)=1+x^2+x^t=(x-1)\Tilde{f}_t(x),\qquad 
        \Tilde{f}_t(x)= x+1 + \sum_{i=0}^{t-1} x^i.
    \end{equation}
    Note that $\Tilde{f}_t(x)$ is nonzero as $\Tilde{f}_t(0)=-1\mod 3$. Because $z_0$ is a root of $f(x^2,x^t)$ and $z_0\ne 1$, it is also a root of $\Tilde{f}_t(x)$. Again, $\Phi_{q^r}(x)$ is a minimal polynomial of $z_0$ in $\Z_3[x]$ so that it must divide $\Tilde{f}_s(x)$. But this is impossible since
    \begin{equation}
        \deg_x \Tilde{f}_t(x)=t-1+\delta_{t,1}\leq t < \varphi(q^r)=\deg_x\Phi_{q^r}(x).
    \end{equation}
    So there is no such $z_0$.
\end{enumerate}
Therefore, when $3$ is a primitive root modulo $q^m$, $(1,1)$ is the solution of \eqref{Eq: variety fsl}. Hence, $\log_3\mathrm{GSD}_\mathrm{FSL}$ is still given by \eqref{Eq: log_3_fsl}.

To conclude, when $L_x=3^{k_x}q^m$ and $L_y=3^{k_y}q^m$ where $k_x,k_y,m\geq 0$ are integers and $q$ is an odd prime such that $p$ is a primitive root modulo $q^m$ (e.g., $q=5$ and any $m\ge 1)$, the ground-state degeneracy of the $\Z_3$ fractal spin liquid is given by
\begin{equation}
    \log_3\mathrm{GSD}_\mathrm{FSL}= 3^{\min(k_x,k_y)}.
\end{equation}

\subsection{$\B$ code of $\Z_3$ Laplacian model on the square lattice}\label{appendix: B sq}

In this case, $q=t=2$, and the stabilizer matrix is given by \eqref{Eq: B_square}
\begin{equation}
    \B^{\Z_3}_\mathrm{sq}=\begin{pmatrix}
        (\B^{\Z_3}_\mathrm{sq})^Z \\
        \hline 0
    \end{pmatrix},\qquad  (\B^{\Z_3}_\mathrm{sq})^Z = \begin{pmatrix}
        -x+y & 1-x \\
        -y+xy & 1-xy
    \end{pmatrix}.
    \label{Eq: appendix_B_square}
\end{equation}
So we use \eqref{Eq: log_p GSD t>1} to compute the GSD for some special values of $L_x$ and $L_y$. The ideal is $\mathfrak i=(f,x^{L_x}-1,y^{L_y}-1)$, where $f(x,y) = -\det (\B^{\Z_3}_\mathrm{sq})^Z = x(y-1)^2+y(x-1)^2 \mod 3$ is precisely the polynomial in \eqref{Eq: anisotropic square Laplacian}\footnote{More precisely, the ideal is $(\bar f, x^{L_x}-1,y^{L_y}-1)$ because $\det ((\B^{\Z_3}_\mathrm{sq})^Z)^\dagger = -\bar f(x,y) = -f(\bar x,\bar y)=-\bar x^2 \bar y^2 f(x,y)$, which is the equivalent to $-f(x,y)$ up to a monomial (i.e., translations).}. Once again, consider the factorization $x^{L_x}-1=(x^{L_x'}-1)^{3^{k_x}} $,  $y^{L_y}-1=(y^{L_y'}-1)^{3^{k_y}} $
and take $L_x'=L_y'=q^m$, where $k_x,k_y,m\ge 0$ are integers and $q$ is an odd prime such that $3$ is a primitive root modulo $q^m$ for $m\ge 1$ (e.g., $q=5$ and any $m\ge 1$). Then, the system of polynomial equations we need to solve is
\begin{equation}
    f(x,y) = x^{q^m}-1 = y^{q^m}-1 = 0.
\end{equation}
In \cite{PhysRevB.107.125121}, it was shown that this system has only the trivial solution $(1,1)$. Therefore, the GSD is given by
\begin{equation}
    \log_3 \mathrm{GSD_{\B,sq}} = \dim_\F \F[x,y]^2/\im \tau,
\end{equation}
where\footnote{The first two columns of this matrix are given by $ x y  \times ((\B^{\Z_3}_\mathrm{sq})^Z)^\dagger$.}
\begin{equation}
    \tau = \begin{pmatrix}
        x-y & 1-x & x^{3^{k_x}}-1 & 0 & y^{3^{k_y}}-1 & 0 \\
        xy-y & xy-1 & 0 & x^{3^{k_x}}-1 & 0 & y^{3^{k_y}}-1
    \end{pmatrix}.
\end{equation}
Assuming without loss of generality that $k_x\ge k_y$, the Gr\"obner basis of $\im \tau$ is given by
\begin{equation}
    \vec g_1 = \begin{pmatrix}
        y^{3^{k_y}}-1\\
        0
    \end{pmatrix},\qquad \vec g_2 = \begin{pmatrix}
        0\\
        y^{3^{k_y}}-1
    \end{pmatrix},\qquad \vec g_3 = \begin{pmatrix}
        x+y+1\\
        y-1
    \end{pmatrix},\qquad \vec g_4 = \begin{pmatrix}
        -y^{3^{k_y}-1}+1\\
        x+y^{3^{k_y}-1}+1
    \end{pmatrix}.
\end{equation}
The set of monomials of $\F[x,y]^2$ that are irreducible with respect to this Gr\"obner basis is
\begin{equation}
    \left\{\begin{pmatrix}
        1\\0
    \end{pmatrix},\ldots,\begin{pmatrix}
        y^{3^{k_y}-1}\\0
    \end{pmatrix},\begin{pmatrix}
        0\\1
    \end{pmatrix},\ldots,\begin{pmatrix}
        0\\y^{3^{k_y}-1}
    \end{pmatrix}\right\},
\end{equation}
with cardinality $2\times 3^{k_y}$. By \eqref{Eq: log_p GSD t>1}, the GSD is
\begin{equation}
    \log_3 \mathrm{GSD_{\B,sq}} = 2 \times 3^{\min(k_x,k_y)}.
\end{equation}

\begin{proof}[Proof of Gr\"obner basis]
    It is easy to check that the above Gr\"obner basis satisfies the Buchberger's criterion. It is also easy to see that
    \begin{equation}
        \begin{aligned}
            &\vec g_3 = \tau_2 - \tau_1,\qquad \vec g_4 = -y^{3^{k_y}-1}(\tau_1+\tau_2) - \tau_5 - (x+1) \tau_6,
            \\
            &\tau_1 = \vec g_1 - \vec g_2 + \vec g_3 + y\vec g_4,\qquad \tau_2 = \vec g_1 - \vec g_2 - \vec g_3 + y\vec g_4.
        \end{aligned}
    \end{equation}
    where $\tau_i$ denotes the $i$-th column of $\tau$. Therefore, $\{\vec g_1,\ldots,\vec g_4\} \in \im \tau$ and $\tau_1,\tau_2 \in \langle \vec g_1,\ldots,\vec g_4 \rangle$. All that is left is to show that $\tau_3,\tau_4\in \langle \vec g_1,\ldots,\vec g_4 \rangle$.

    It is convenient to work with variables $u=x-1$ and $v=y-1$ with monomial order $u\succ v$. Then the matrix $\tau$ becomes
    \begin{equation}
        \tilde\tau = \begin{pmatrix}
            u-v & -u & u^{3^{k_x}} & 0 & v^{3^{k_y}} & 0 \\
            uv+u & uv+u+v & 0 & u^{3^{k_x}} & 0 & v^{3^{k_y}}
        \end{pmatrix},
    \end{equation}
    and the Gr\"obner basis is
    \begin{equation}
        \tilde{\vec g}_1 = \begin{pmatrix}
            v^{3^{k_y}}\\
            0
        \end{pmatrix},\qquad \tilde{\vec g}_2 = \begin{pmatrix}
            0\\
            v^{3^{k_y}}
        \end{pmatrix},\qquad \tilde{\vec g}_3 = \begin{pmatrix}
            u+v\\
            v
        \end{pmatrix},\qquad \tilde{\vec g}_4 = \begin{pmatrix}
            v\sum_{i=0}^{3^{k_y}-2} (-v)^i\\
            u-v\sum_{i=0}^{3^{k_y}-2} (-v)^i
        \end{pmatrix}.
    \end{equation}
    Consider a general vector of the form
    \begin{equation}
        \vec s = \sum_{i=0}^k u^{k-i} v^i \begin{pmatrix}
            s_{1,i}(v)\\
            s_{2,i}(v)
        \end{pmatrix},
    \end{equation}
    where $s_{1,i}(v)$ and $s_{2,i}(v)$ are polynomials in $v$ independent of $u$. For example, $\tilde \tau_3$, $\tilde \tau_4$, $\tilde{\vec g}_3$, and $\tilde{\vec g}_4$ are all of this form. We want to reduce $\vec s$ with respect to $\tilde{\vec g}_3$ and $\tilde{\vec g}_4$ as much as possible. First, we have
    \begin{equation}
        \vec s^{(1)} = \vec s - s_{1,0}(v) u^{k-1} \tilde{\vec g}_3 - s_{2,0}(v) u^{k-1} \tilde{\vec g}_4.
    \end{equation}
    It is easy to see that $\vec s^{(1)}$ is of the same form as $\vec s$ and the leading terms of both components of $\vec s^{(1)}$ are smaller than the leading terms of the corresponding components of $\vec s$. Repeating this $k$ times, we end up with a vector of the form
    \begin{equation}
        \vec s^{(k)} = v^k \begin{pmatrix}
            s^{(k)}_{1,k}(v)\\
            s^{(k)}_{2,k}(v)
        \end{pmatrix},
    \end{equation}
    which is not reducible with respect to $\tilde{\vec g}_3$ or $\tilde{\vec g}_4$ anymore.

    Applying this procedure to $\tilde \tau_3$ and $\tilde \tau_4$, we end up with a vector proportional to $v^{3^{k_x}}$, which is reducible with respect to $\tilde{\vec g}_1$ and $\tilde{\vec g}_2$ because $k_x \ge k_y$. Therefore, $\tilde\tau_3,\tilde\tau_4\in \langle \tilde{\vec g}_1,\ldots,\tilde{\vec g}_4 \rangle$.
\end{proof}

~
\twocolumngrid

\section{Mobility of $z$-lineons in the triangular-based anisotropic Laplacian model\label{appendix: mobility_tri_laplacian}}
As mentioned in the main text, the point-like excitations of the anisotropic Laplacian model can always move in the $z$ direction, making them $z$-lineons. Their mobility in the $xy$ plan is quite nontrivial and can be analyzed in the polynomial formalism \cite{PhysRevB.94.235157}. Consider a configuration of $Q$ $z$-lineons with positions $(n_i,m_i)$, described by the Laurent polynomial (i.e., an element of $\Z_N[x,x^{-1},y,y^{-1}]$),
\begin{equation}
    q(x,y)=\sum_{i=1}^Q x^{n_i}y^{m_i}.
\end{equation}
Assume this configuration of $Q$ lineons can rigidly move to another position $(n_0,m_0)\neq (0,0)$. This is equivalent to the existence of a Laurent polynomial $s(x,y)$ such that \cite{PhysRevB.107.125121,PhysRevB.108.075106}
\begin{equation}
    (x^{n_0}y^{m_0}-1)q(x,y)=s(x,y)f(x,y) \mod N.
    \label{Eq: mobility_condition}
\end{equation}
Here, $f(x,y)=x^2+y^2+x^2y+xy^2+x+y-6xy$ is the stabilizer polynomial given by the top left entry of \eqref{Eq: tri_matrix}. Physically, \eqref{Eq: mobility_condition} means that the configurations at the initial position $(0,0)$ and the final position $(n_0,m_0)$ can be connected by the Hamiltonian terms (trivial charge configurations) $f(x,y)$ without creating additional excitations. 

If $q(x,y)$ can be written as $q(x,y)=r(x,y)f(x,y)$ for some Laurent polynomial $r(x,y)=\sum_j x^{n'_j}y^{m'_j}$, then the condition \eqref{Eq: mobility_condition} is trivially satisfied by choosing $s(x,y)=(x^{n_0}y^{m_0}-1)r(x,y)$. A more interesting case is when $q(x,y)\neq r(x,y)f(x,y)$ for any $r(x,y)$, which means $f(x,y)$ and $x^{n_0}y^{m_0}-1$ must share a nontrivial factor. 

When $N=2$, the stabilizer polynomial can be factorized into $f(x,y)=(x+y)(1+x)(1+y) \mod 2$. Taking $(n_0,m_0)=(n,-n)$, $x^ny^{-n}+1$ can be factorized as
\begin{equation}
    (x+y)(x^{n-1}+x^{n-2}y+\cdots+y^{n-1})\mod 2,
\end{equation}
up to a monomial $y^{-n}$. The condition \eqref{Eq: mobility_condition} is thus satisfied when $q(x,y)=(1+x)(1+y) =1+x+y+xy$. That is, a quadrupole of $z$-lineons located at the relative positions $(0,0)$, $(1,0)$, $(0,1)$ and $(1,1)$ can move along the $(\pm 1,\mp 1)$ direction, as illustrated in Fig. \ref{fig: tri_mobility}. Using the above factorization of $f(x,y)$, we can further show that there are two other mobile configurations: a quadrupole of $(0,0)$, $(1,0)$, $(0,1)$, $(-1,1)$, and that of $(0,0)$, $(0,1)$, $(-1,1)$, $(-1,2)$ can move in the $(0,\pm 1)$ and $(\pm 1,0)$ directions, respectively.

When $N = p$ is an odd prime, we show that the condition \eqref{Eq: mobility_condition} cannot be satisfied, except trivially as in $q(x,y) = r(x,y) f(x,y)$. Let $p^k$ be the largest power of $p$ that divides $n_0$ and $m_0$, i.e., $n_0'=n_0/p^k$, $m_0'=m_0/p^k$, and $d=\gcd(n_0',m_0')$ is not divisible by $p$. First we observe that $x^{n_0}y^{m_0}-1$ can be factorized as
\begin{equation}
    (x^{n'_0}y^{m'_0}-1)^{p^k}=[(x^{n''_0}y^{m''_0}-1)t(x,y)]^{p^k} \mod p,
\end{equation}
where $n_0''=n_0'/d$, $m_0''=m_0'/d$ and $t(x,y)=\sum_{i=0}^{d-1}(x^{n_0''}y^{m_0''})^i $. 
Specifically, for $N=3$, the stabilizer polynomial $f(x,y)$ is factorizable as $(1+x+y)(x+y+xy)\mod 3$. All we need to show is that $x^{n_0}y^{m_0}-1 $ is not a multiple of $f_1(x,y)=1+x+y$ or $f_2(x,y)=x+y+xy$. Note that $t(1,1)=d\neq 0 \mod 3$ but $f_1(1,1)=f_2(1,1)=0 \mod 3$. Therefore, $t(x,y)^{p^k}$ is not a multiple of $f_1(x,y)$ and $f_2(x,y)$. Also $x^{n''_0}y^{m''_0}-1$ is irreducible up to a monomial so $(x^{n''_0}y^{m''_0}-1)^{p^k} $ is also not a multiple of either of them. Therefore, $x^{n_0}y^{m_0}-1$ is not a multiple of $f(x,y)$ for $N=3$. When $N=p>3$, the stabilizer polynomial $f(x,y)$ is irreducible up to a monomial.\footnote{For $p>5$, the irreducibility of $f(x,y)$ modulo $p$ can be verified using \cite[Corollary 3]{GAO200332}. For $p=5$, this can be done in a computer algebra system, such as Mathematica.} Using the same argument again, we can show that $x^{n_0}y^{m_0}-1$ is not a multiple of $f(x,y)$. 

In conclusion, we find that any finite set of $z$-lineons in the triangular-based anisotropic $\Z_p$ Laplacian model with odd prime $p$ cannot move in the $xy$ plane, except when they are created locally.

\section{Robustness of the anisotropic Laplacian model on various regular lattices}\label{appendix: robust-kagome}
In this appendix, we prove the robustness of the anisotropic $\mathbb Z_N$ Laplacian model on the extended triangular, honeycomb, and Kagome lattices. We will show that, on the Kagome lattice, robustness requires $N$ to be coprime to $6$, but there are no such restrictions on $N$ on the triangular and honeycomb lattices.

As explained in Sec. \ref{subsection: anisotropic laplacian triangular}, the logical operators $W_z$ and $\tilde W_z$ have non-local support, so let us focus on $W(h)$ and $\tilde W(h)$, whose support is the same as that of the discrete harmonic function $h$.

\begin{figure}
    \centering
    \includegraphics[scale=0.8]{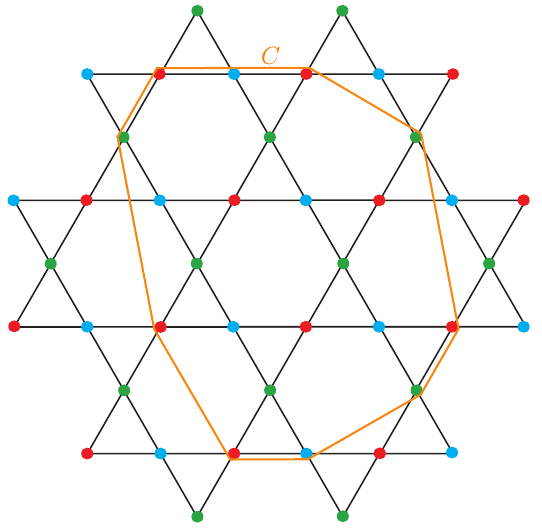}
    \caption{A simple closed convex polygonal curve $C$ whose corners are on the sites of the Kagome lattice. We enlarged the curve a bit so that the sites and links along the boundary are visible.}
    \label{fig: Kagome_robustness_proof}
\end{figure}

Assume $h$ has finite support. Consider a simple closed convex polygonal curve $C$ whose corners are on the lattice sites and whose interior contains the support of $h$ (points on $C$ are considered to be inside $C$). An example curve on Kagome lattice is sketched in Fig. \ref{fig: Kagome_robustness_proof}. It is easy to see that every corner of $C$ has at least two neighbouring sites outside $C$ (a.k.a. outer neighbours). Pick a corner $\vec i$ of $C$. We have two cases.
\begin{enumerate}
    \item There is an outer neighbour of $\vec i$, call it $\vec j$, such that the remaining three neighbours of $\vec j$ are all outside $C$. (This is always the case on the triangular and honeycomb lattices, but not on the Kagome lattice---e.g., this is the case for all the corners of $C$ in Fig. \ref{fig: Kagome_robustness_proof} except for the red and green corners on the top-left.) In this case, using the discrete Laplacian equation $\Delta_L h(\vec j) = 0 \mod N$ at the site $\vec j$, we find that $h(\vec i) = 0 \mod N$. So we can shrink $C$ inward to get a curve with smaller interior but still contains the support of $h$.
    
    \item On the Kagome lattice, if the condition of the last case is not satisfied, then there are two outer neighbours of $\vec i$, call them $\vec j_1$ and $\vec j_2$, such that $\vec j_1$ (resp. $\vec j_2$) has another neighbour $\vec i_1$ (resp. $\vec i_2$) inside $C$. (This is the case, for example, for the red and green corners on the top-left of $C$ in Fig. \ref{fig: Kagome_robustness_proof}.) In this case, using the discrete Laplacian equation at the sites $\vec j_1$ and $\vec j_2$, we get $h(\vec i) = - h(\vec i_1) = - h(\vec i_2) \mod N$. Combining these equations with the discrete Laplacian equation at the site $\vec i$, we find that $6h(\vec i) = 0 \mod N$. When $N$ is coprime to $6$, we get $h(\vec i) = 0 \mod N$, which also implies that $h(\vec i_1) = h(\vec i_2) = 0 \mod N$. Therefore, once again, we can shrink $C$ inward to get a curve with smaller interior but still contains the support of $h$.
\end{enumerate}
Repeating these steps, by induction, we conclude that $h=0$ everywhere. Hence, the anisotropic $\Z_N$ Laplacian model on triangular and honeycomb lattices is robust for all $N$, whereas it is robust on the Kagome lattice if and only if $N$ is coprime to $6$. (Recall that, in Sec. \ref{subsection: anisotropic Laplacian_Kagome}, we constructed local logical operators on the Kagome lattice when $N$ is a multiple of $2$ or $3$.)

\bibliography{refs}
\end{document}